\documentclass[prb,twocolumn,aps,showpacs]{revtex4}
\usepackage{graphicx}%
\usepackage{dcolumn}
\usepackage{amsmath}
\usepackage{amssymb}
\usepackage{epsfig}

\begin{document}
\title{Low-energy theory of the  Nambu-Goldstone modes  of an ultracold $^6Li-$ $^{40}K$ mixture in an optical lattice }
\author{Z. G. Koinov}\affiliation{Department of Physics and Astronomy,
University of Texas at San Antonio, San Antonio, TX 78249, USA}
\email{Zlatko.Koinov@utsa.edu}
 \begin{abstract}
 A low-energy theory of the  Nambu-Goldstone  excitation spectrum and the corresponding speed of sound of  an  interacting Fermi mixture of Lithium-6 and
Potassium-40 atoms in a two-dimensional optical lattice at finite
temperatures  with the Fulde-Ferrell order parameter has been formulated. It is assumed that the two-species interacting Fermi gas is described by the one-band Hubbard Hamiltonian  with an attractive on-site interaction. The discussion is restricted to the  BCS side of the Feshbach resonance where the  Fermi atoms exhibit superfluidity.  The quartic on-site interaction is decoupled via a Hubbard-Stratonovich  transformation by introducing a four-component boson field which mediates the Hubbard interaction. A functional integral
  technique and a Legendre transform  are used to give a systematic derivation of the Schwinger-Dyson equations for the generalized single-particle Green's function and the Bethe-Salpeter equation for the two-particle Green's function and the associated collective modes. The numerical solution of the Bethe-Salpeter equation  in the generalized random phase approximation shows that there exist two distinct sound
velocities in the long-wavelength limit. In addition to the
long-wavelength mode (Goldstone mode), the two-species  Fermi gas has a superfluid phase revealed by two rotonlike minima in the asymmetric collective-mode energy.
\end{abstract}
 \maketitle
\section{Introduction}
Optical lattices are formed by the interference of counter
propagating laser beams. If the laser beams have equal frequencies,
the gases of ultracold alkali atoms can be trapped in periodic
potentials  created by standing waves of laser light.
Because of the Stark effect the ground-state alkali atoms couple to
the electromagnetic field
 via an induced electric dipole moment. From theoretical point of view, the simplest approach to the trapped fermions is the tight-binding approximation,
 which requires  sufficiently deep lattice potential. In the tight-binding limit, two alkali atoms of opposite
pseudospins on the same site have  an interaction energy $U$, while
the probability to tunnel to a neighboring site is given by the
hopping parameters. The hopping parameters as well as the interaction
energy depend on the depth of the  lattice potential  and can be
 tuned  by varying the intensity of the laser beams. We  assume that the interacting
fermions are in a sufficiently deep periodic lattice potential
described by the Hubbard Hamiltonian. We  restrict the discussion to
the case of atoms confined to the lowest-energy band (single-band
Hubbard model), with two possible  states described by
pseudospins $\sigma$.  We consider different amounts of $^6Li$ and $^{40}K$ atoms in each  state ($\sigma=\uparrow=Li$, $\sigma=\downarrow=K$)
achieved by considering different chemical potentials $\mu_\uparrow$ and $\mu_\downarrow$. We assume that
there are $M = M_\uparrowª+M_\downarrow$ atoms distributed along $N$ sites, and the corresponding filling
factors $f_{\uparrow,\downarrow}=M_{\uparrow,\downarrow}/N$ are smaller than unity. The Hubbard Hamiltonian is defined as follows:
\begin{equation}H=-\sum_{<i,j>,\sigma}J_{\sigma}\psi^\dag_{i,\sigma}\psi_{j,\sigma}
-U\sum_i \widehat{n}_{i,\uparrow}
\widehat{n}_{i,\downarrow}-\sum_{i,\sigma}\mu_\sigma\widehat{n}_{i,\sigma},\label{H}\end{equation}
where $J_{\sigma}$ is the single electron hopping integral,  and
 $\widehat{n}_{i,\sigma}=\psi^\dag_{i,\sigma}\psi_{i,\sigma}$ is the
density operator on site $i$. The Fermi operator
$\psi^\dag_{i,\sigma}$ ($\psi_{i,\sigma}$) creates (destroys) a
fermion on the lattice site $i$  with pseudospin projection
$\sigma$. The symbol $\sum_{<ij>}$ means sum over nearest-neighbor
sites of the two-dimensional lattice. The first term in (\ref{H}) is the usual kinetic
energy term in a tight-binding approximation. All numerical calculations will be performed  assuming that
 the hopping (tunneling) ratio $J_{Li}/J_K \approx 0.15$. In our notation  the strength of the on-site interaction $U>0$ is positive, but the
negative sign in front of the interaction corresponds to the Hubbard
model with an attractive interaction. In the presence of an
(effective) attractive interaction between the fermions, no matter
how weak it is, the alkali atoms form bound pairs, also called
the Cooper pairs. As a result, the system becomes unstable against the
formation of a new many-body  superfluid ground state. The superfluid ground state comes from the U(1) symmetry breaking and it is
characterized by a nonzero order parameter, which in the population-balanced case is assumed to be
a constant in space $\Delta_0$. Physically, it describes superfluid
state of  Cooper pairs with zero momentum. Superfluid state of
Cooper pairs with nonzero momentum  occurs in population-imbalanced case between a fermion with momentum
$\textbf{k} + \textbf{q}$ and spin $\uparrow$ and a fermion with
momentum $-\textbf{k} + \textbf{q}$, and spin $\downarrow$ . As a
result, the pair momentum is $2\textbf{q}$. A finite pairing momentum implies a position-dependent phase of the order parameter, which in the  Fulde-Ferrell\cite{FF} (FF) case  varies as a single plane wave
$\Delta(\textbf{r})=\Delta_\textbf{q}\exp\left(2\imath
\textbf{q.r}\right)$, where $\Delta_\textbf{q}$ is a real quantity. The order parameter also can be a combination of two plane waves as in the case of the Larkin-Ovchinnikov\cite{LO} (LO) superfluid states.   In both cases we are dealing with a spontaneous translational symmetry breaking and with an inhomogeneous superfluid state. When
continuous and global symmetries are spontaneously broken the collective modes known as the Nambu-Goldstone\cite{N,G} (NG) modes appear. From experimental point of view, the NG dispersion can be measured with unprecedented precision in systems of ultracold fermionic atoms on an optical lattice.\cite{SFexp1,SFexp2,SFexp3,SFexp4,SFexp5,SFexp6,SFexp7,SFexp8}

Turning our attention to the theoretical description of the single-particle  and collective-mode excitations of superfluid alkali atom Fermi gases in
optical lattice potentials, we find that there have been  impressive theoretical achievements.
\cite{SF0,SF1,SF1a,SF2,SF3,SF4,SF5,SF6,SF7,SF8,SF9,SF9a,SF9b,SF10,SF11,SF12,SF13,SF14,SF15,SF15a,SF16,SF16a,SF17,SF17a,SF18,
SF19,SF19a,SF19b,SF19c,SF19d,SF19e,SF19g,SF20,SF21,SF22}
Generally speaking, the single-particle excitations manifest themselves as poles of the single-particle
Green's function, $G$; while the two-particle (collective)
excitations could be related to the poles of the  two-particle
Green's function, $K$. The poles of these Green's functions are
defined by the solutions of the Schwinger-Dyson (SD) equation\cite{Schwinger,Dyson}
$G^{-1}=G^{(0)-1}-\Sigma$,  and the Bethe-Salpeter (BS) equation\cite{BetheS}
$[K^{(0)-1}-I]\Psi=0$, respectively. Here,  $G^{(0)}$ is the
 free single-particle propagator, $\Sigma$ is the fermion self-energy,
$I$ is the BS kernel, and the two-particle free propagator $K^{(0)}
= GG $ is a product of two fully dressed single-particle Green's
functions.  Since the fermion self-energy
 depends on the two-particle Green's function, the
positions of both poles must to be obtained by solving
 the SD and BS equations self-consistently.

 Instead of solving the SD and BS equations self-consistently, it is widely accepted that the single-particle dispersion can be  obtained in the mean-field approximation or by solving the Bogoliubov-de Gennes (BdG) equations in a self-consistent fashion, while  the generalized random phase approximation
(GRPA) is the one that can provide the collective excitations in a
weak-coupling regime.  In the GRPA, the single-particle
excitations are replaced with those obtained by diagonalizing the
Hartree-Fock (HF) Hamiltonian; while the collective modes are
obtained by solving the BS equation in which the single-particle
Green's functions are calculated in HF approximation, and the BS
kernel is obtained by summing ladder and bubble diagrams.

There exist two different formulations of the
GRPA that can be used to calculate the spectrum of the collective
excitations of the Hubbard Hamiltonian (\ref{H}). The first approach
uses the Green's function
method,\cite{SF16a,CGexc,CG1,CCexc,ZKexc,Com,ZGK,ZK1}  while the
second one is based on the Anderson-Rickayzen
equations.\cite{PA,R,BR,SF17}

The Green's function approach has been used to obtain the
collective excitations in the problems of the Bose-Einstein condensation (BEC) of excitons (or
excitonic polaritons) in semiconductors,\cite{CGexc,CCexc,ZKexc} and
the BEC of Cooper pairs in s-wave layered superconductors.
\cite{CG1} According to the Green's function method,  the collective
modes manifest themselves as poles of  the two-particle Green's
function, $K$, as well as the density and spin response functions.
Both response functions can be expressed in terms of  $K$, but it is
very common to obtain the poles of the density response function by
following the Baym and Kadanoff formalism,\cite{BK,BK2} in which the
density response function is defined in terms of functional
derivatives of the density, with respect to the external
fields.\cite{SF16a,SF19}

The second method that can be used to obtain the collective
excitation spectrum of the Hubbard Hamiltonian  starts from the
Anderson-Rickayzen equations, which in the GRPA can be reduced to a
set of three coupled equations, such that the collective-mode
spectrum is obtained by solving  a $3\times 3$ secular
determinant.\cite{BR,SF17}

From theoretical point of view, the corresponding expressions for the Green's functions cannot be
evaluated exactly because the  interaction part
of the Hubbard Hamiltonian is quartic in the fermion fields. The simplest way to solve this problem is to
apply the so-called mean-field decoupling of the quartic
interaction. To go beyond the mean-field approximation, we apply the
idea that we can transform the quartic term
 into  quadratic form by making the
Hubbard-Stratonovich  transformation  for the fermion operators. In
contrast to the previous approaches, such that  after performing the
Hubbard-Stratonovich  transformation the fermion degrees of freedom
are integrated out; we decouple the quartic problem by introducing a
model system which consists of a multi-component boson field
$A_\alpha$ interacting with  fermion fields $\psi^\dagger$ and
$\psi$.

 The functional-integral formulation of the Hubbard
model requires the representation of the Hubbard interaction
of (\ref{H}) in terms of squares of one-body charge
and spin operators. It is known that it may be possible to
resolve the Hubbard interaction into quadratic forms of
spin and electron number operators in an infinite
number of ways.\cite{ZR} If no approximations were made in evaluating the functional integrals, it would no matter which of the ways is chosen. When approximations are taken, the final result depends on a particular form chosen.
Thus, one should check that the results obtained with the Hubbard-Stratonovich  transformation are consistent with the results obtained with the canonical mean-field
approximation. It can be seen that our approach to the Hubbard-Stratonovich transformation provides  results consistent with the results obtained with the mean-field
approximation, i.e. one can derive the mean-field gap equation using the collective-mode dispersion $\omega(Q)$ in the limit $Q\rightarrow 0$ and $\omega\rightarrow 0$.

 There are three
advantages of keeping both the fermion and the boson degrees of
freedom. First,  the approximation that is used to decouple the
self-consistent relation between the fermion self-energy and the
two-particle Green's function automatically leads to conserving
approximations because it relies on the fact that the BS kernel
 can be written as functional derivatives of the Fock
$\Sigma^F$ and the Hartree $\Sigma^H$ self-energy
$I=I_d+I_{exc}=\delta \Sigma^F/\delta G+\delta \Sigma^H/\delta
G=\delta^2 \Phi/\delta G\delta G$. As shown by Baym, \cite{BK2} any
self-energy approximation is conserving whenever: (i) the
self-energy  can be written as the derivative of a functional
$\Phi[G]$, i.e. $\Sigma=\delta \Phi[G]/\delta G$, and (ii) the SD
equation for $G$ needs to be solved fully self-consistently for this
form of the self-energy.   Second, the
 collective excitations of the Hubbard model can be calculated in two different ways: as poles of
  the fermion Green's function, $K$, and  as  poles
  of the boson Green's function, $D$; or equivalently,  as poles of the density and spin
parts of the general response function, $\Pi$. Here, the boson Green's
function, $D$, is defined by the Dyson equation
  $D=D^{(0)}+D^{(0)}\Pi D^{(0)}$ where $D^{(0)}$ is the free
boson propagator. Third, the action which describes the interactions
in the Hubbard  model is similar to the action $\psi^\dagger A\psi$
in quantum electrodynamics. This allows us to apply  powerful
field-theoretical methods, such as the method of Legendre
transforms,\cite{DM} to derive the SD and BS equations, as well as
the
 vertex equation for the vertex function, $\Gamma$,
 and the Dyson equation for the boson Green's function, $D$.

 \begin{figure}\includegraphics[scale=0.8]{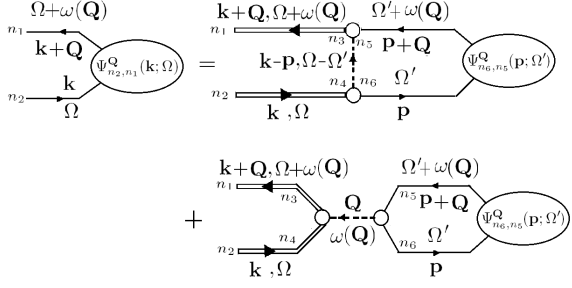}
    \caption{Diagrammatic representation of the Bethe-Salpeter equation (\ref{BSEdZ}) for the BS amplitude
    $\Psi^{\textbf{Q}}_{n_2,n_1}(\textbf{k};\Omega)$. The single-particle Green's function $G_{n_2n_1}(\textbf{k},\Omega)$
     is represented by two solid lines, oriented in the direction of fermion propagation. The dashed lines
      represent the free boson propagator $D^{(0)}_{\alpha\beta}(\omega,\textbf{Q})$.
      At each of the bare vertices $\Gamma_\alpha^{(0)}(n_1,n_2)$ (represented by  circles)  the energy and momentum are conserved.}\end{figure}
 The basic assumption in the BS formalism  is that the
bound states of two fermions in the optical lattice at zero
temperature  are described by the BS wave functions (BS amplitudes).
The BS amplitude determines the probability amplitude to find the
first fermion  at the site $i$ at the moment $t_1$ and the second
fermion at the site $j$ at the moment $t_2$. The BS amplitude
depends on the relative internal time $t_1-t_2$ and on the
"center-of-mass" time $(t_1+t_2)/2$:\cite{IZ}
\begin{equation}\begin{split}&\Phi^{\textbf{Q}}_{n_2n_1}(\textbf{r}_i,\textbf{r}_j;t_1,t_2)=\\&
\exp\left\{\imath\left[\textbf{Q.}(\textbf{r}_i+\textbf{r}_j)/2-\omega(\textbf{Q})(t_1+t_2)/2\right]\right\}\\&
\Psi^\textbf{Q}_{n_2n_1}(\textbf{r}_i-\textbf{r}_j,t_1-t_2),\end{split}\end{equation}
where $\{n_1, n_2\}$ are the corresponding quantum numbers,
$\textbf{Q}$ and $\omega(\textbf{Q})$ are the collective-mode
momentum, and its dispersion, respectively.  The Fourier transform $\Psi^{\textbf{Q}}_{n_2,n_1}(\textbf{k};\Omega)$
of the BS amplitude
$$\Psi^\textbf{Q}_{n_2n_1}(\textbf{r},t)=\int
\frac{d\Omega}{2\pi}\int\frac{d^d\textbf{k}}{(2\pi)^d}e^{\imath\left(\textbf{k}\textbf{.r}-\Omega
t\right)}\Psi^{\textbf{Q}}_{n_2,n_1}(\textbf{k};\Omega)$$ satisfies
the BS equation,  presented diagrammatically in Fig. 1.
The direct interaction in the case of
the Hubbard  model  is frequency independent;
therefore the following BS equation for the equal-time BS amplitude
$\Psi^{\textbf{Q}}_{n_2,n_1}(\textbf{k})=\int
\frac{d\Omega}{2\pi}\Psi^{\textbf{Q}}_{n_2,n_1}(\textbf{k};\Omega)$
takes place:
\begin{equation}\begin{split}
&\Psi^{\textbf{Q}}_{n_2,n_1}(\textbf{k})=\int \frac{d\Omega}{2\pi}
G_{n_1n_3}\left(\textbf{k}+\textbf{Q},\Omega+\omega(\textbf{Q})\right)G_{n_4n_2}(\textbf{k},\Omega)\\&\int\frac{d^d\textbf{p}}{(2\pi)^d}
\left[I_d\left(%
\begin{array}{cc}
  n_3 & n_5  \\
  n_4 & n_6 \\
\end{array}|\textbf{k}-\textbf{p}
\right)+I_{exc}\left(%
\begin{array}{cc}
  n_3 & n_5  \\
  n_4 & n_6 \\
\end{array}|\textbf{Q}
\right)\right]\times\\&\Psi^{\textbf{Q}}_{n_6,n_5}(\textbf{p}).\label{BSEdZ}
\end{split}\end{equation}
Here, $G_{n_1n_2}\left(\textbf{k},\Omega\right)$ is the Fourier transform of the
single-particle dressed Green's function, and $I_d$ and $I_{exc}$ are the
direct and exchange parts of the BS kernel defined as functional derivatives of the Fock
and the Hartree self-energies.

The superfluid states can be described in terms of the Namby-Gor'kov single-particle  Green's function which is a thermodynamic average of the
$\widehat{T}_u$-ordered tensor product of  the Nambu field operators
 \begin{equation}\widehat{\overline{\psi}}(y)
=\left(\psi^\dag_{\uparrow}(y)
\psi_{\downarrow}(y) \right),\quad \widehat{\psi}(x)=\left(%
\begin{array}{c}
  \psi_{\uparrow}(x)  \\
  \psi^\dag_{\downarrow}(x) \\
\end{array}%
\right)\label{NGFO}.\end{equation}
As suggested by Maki,\cite{KM}  it is more convenient to use a spinor representation of the single-particle state by introducing the following four-component fermion fields \begin{equation}\begin{split}&\widehat{\Psi}(x)=\left(%
\begin{array}{c}
  \psi_\uparrow(x) \\
  \psi_\downarrow(x) \\
  \psi^\dag_\uparrow(x)\\
  \psi^\dag_\downarrow(x)\\
\end{array}%
\right),\\&\widehat{\Psi}^\dag
(y)=\left(\psi^\dag_\uparrow(y)\psi^\dag_\downarrow(y)
\psi_\uparrow(y)\psi_\downarrow(y)
\right).\label{NG1}\end{split}\end{equation}
Here, we introduce composite variables
$y=\{\textbf{r}_i,u\}=\{i,u\}$ and
$x=\{\textbf{r}_{i'},u'\}=\{i',u'\}$, where
$\textbf{r}_{i},\textbf{r}_{i'}$ are the lattice
site vectors, and according to imaginary-time (Matsubara) formalism
the variable $u,u'$ range from $0$ to $\hbar \beta=\hbar/(k_BT)$. Throughout
this paper we have assumed $\hbar=k_B=1$, the lattice constant $a=1$, and we use the
summation-integration convention: that repeated variables are summed
up or integrated
over.

The field operators (\ref{NGFO}) allow us to define the Nambu-Gor'kov single-particle Green's function:
\begin{equation}\begin{split}&\widehat{G}(x_1;y_2)
=-<\widehat{T}_u\left(\widehat{\psi}(x_1)\otimes\widehat{\overline{\psi}}(y_2)\right)>= \\&-\left(%
\begin{array}{cc}
  <\widehat{T}_u\left(\psi_\uparrow(x_1)\psi^\dag_\uparrow(y_2)\right)> &
 <\widehat{T}_u\left(\psi_\uparrow(x_1)\psi_\downarrow(y_2)\right)> \\
     <\widehat{T}_u\left(\psi^\dag_\downarrow(x_1)\psi^\dag_\uparrow(y_2)\right)>
      &
   <\widehat{T}_u\left(\psi^\dag_\downarrow(x_1)\psi_\downarrow(y_2)\right)>  \\
\end{array}%
\right),\label{G2}\end{split}
\end{equation}
and the generalized single-particle Green's function which includes all possible   thermodynamic averages:
\begin{widetext}
\begin{eqnarray}&\widehat{G}(x_1;y_2)=-<\widehat{T}_u\left(\widehat{\Psi}(x_1)\otimes\widehat{\overline{\Psi}}(y_2)\right)>\nonumber\\& -\left(%
\begin{array}{cccc}
  <\widehat{T}_u\left(\psi_\uparrow(x_1)\psi^\dag_\uparrow(y_2)\right)> &
<\widehat{T}_u\left(\psi_\uparrow(x_1)\psi^\dag_\downarrow(y_2)\right)>
   & <\widehat{T}_u\left(\psi_\uparrow(x_1)\psi_\uparrow(y_2)\right)> &
     <\widehat{T}_u\left(\psi_\uparrow(x_1)\psi_\downarrow(y_2)\right)> \\
  <\widehat{T}_u\left(\psi_\downarrow(x_1)\psi^\dag_\uparrow(y_2)\right)>
   &
   <\widehat{T}_u\left(\psi_\downarrow(x_1)\psi^\dag_\downarrow(y_2)\right)>
   & <\widehat{T}_u\left(\psi_\downarrow(x_1)\psi_\uparrow(y_2)\right)>&
     <\widehat{T}_u\left(\psi_\downarrow(x_1)\psi_\downarrow(y_2)\right)>\\
   <\widehat{T}_u\left(\psi^\dag_\uparrow(x_1)\psi^\dag_\uparrow(y_2)\right)>
   &
   <\widehat{T}_u\left(\psi^\dag_\uparrow(x_1)\psi^\dag_\downarrow(y_2)\right)>

   & <\widehat{T}_u\left(\psi^\dag_\uparrow(x_1)\psi_\uparrow(y_2)\right)>&
     <\widehat{T}_u\left(\psi^\dag_\uparrow(x_1)\psi_\downarrow(y_2)\right)>\\
   <\widehat{T}_u\left(\psi^\dag_\downarrow(x_1)\psi^\dag_\uparrow(y_2)\right)>
   &<\widehat{T}_u\left(\psi^\dag_\downarrow(x_1)\psi^\dag_\downarrow(y_2)\right)>
   & <\widehat{T}_u\left(\psi^\dag_\downarrow(x_1)\psi_\uparrow(y_2)\right)>
   &
   <\widehat{T}_u\left(\psi^\dag_\downarrow(x_1)\psi_\downarrow(y_2)\right)>  \\
\end{array}%
\right). \label{EGF}
\end{eqnarray}
In the tight-binding approximation the BS equation  (in the GRPA) can be reduced to a secular determinant, which determines the collective-mode dispersion. The  Nambu-Gor'kov single-particle Green's function (\ref{G2})   leads  to the following $4\times 4$ secular determinant:\cite{SF18}
\begin{equation}
Z_{4}=\left|
\begin{array}{cccc}
U^{-1}+\left(I_{\gamma,\gamma}-L_{\widetilde{\gamma},\widetilde{\gamma}}\right)&\left(J_{\gamma,l}-K_{m,\widetilde{\gamma}}\right)&
\left(I_{\gamma,\widetilde{\gamma}}+L_{\gamma,\widetilde{\gamma}}\right)&\left(J_{\gamma,m}+K_{l,\widetilde{\gamma}}\right)\\
\left(J_{\gamma,l}-K_{m,\widetilde{\gamma}}\right)&U^{-1}+\left(I_{l,l}-L_{m,m}\right)&
\left(J_{l,\widetilde{\gamma}}+K_{m,\gamma}\right)&\left(I_{l,m}+L_{l,m}\right)\\
\left(I_{\gamma,\widetilde{\gamma}}+L_{\gamma,\widetilde{\gamma}}\right)&\left(J_{l,\widetilde{\gamma}}+K_{m,\gamma}\right)&
-U^{-1}+\left(I_{\widetilde{\gamma},\widetilde{\gamma}}-L_{\gamma,\gamma}\right)&\left(J_{\widetilde{\gamma},m}-K_{\gamma,l}\right)\\
\left(J_{\gamma,m}+K_{l,\widetilde{\gamma}}\right)&\left(I_{l,m}+L_{l,m}\right)&
\left(J_{\widetilde{\gamma},m}-K_{\gamma,l}\right)&U^{-1}+\left(I_{m,m}-L_{l,l}\right)
\end{array}%
\right|.\label{SecDet4}\end{equation}
\end{widetext}
 It is well known  that when the  generalized Green's function (\ref{EGF}) is used, the BS approach  should provide a $16 \times 16$ secular determinant. In what follows, we  derive the corresponding  $16 \times 16$ secular determinant for a system  of a $^6Li-$ $^{40}K$ mixture loaded
in a two-dimensional  optical lattice, and by means of it, we  calculate numerically the collective-mode dispersion.

\begin{figure} \includegraphics[scale=0.75]{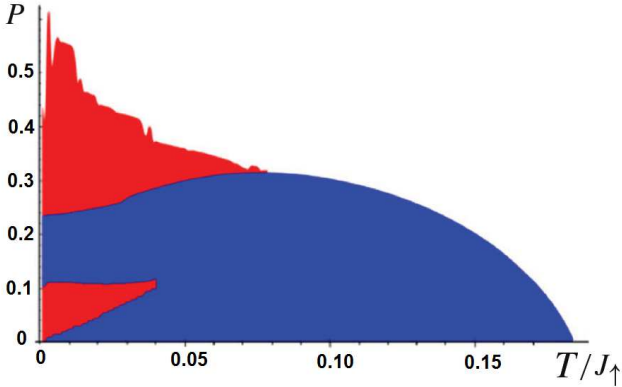}
    \caption{The phase diagrams of a $^6Li-$$^{40}K$ mixture in a square lattice.\cite{SZ} The interaction strength is
$U = 2J_\uparrow$.  The polarization is defined as $P = ( f_K - f_{Li} )/ f $,
where the total filling is $f = 0.5$ atoms/lattice site. Colors: Sarma states = blue (black), FF = red
(dark grey), and normal gas = white.   }\end{figure}

 The mean-field treatment of the FF and LO
phases in a variety of systems, such as superconductors with Zeeman
splitting, heavy-fermion
superconductors,\cite{SC1,SC2,SC3,SC4,SC5,SC6,SC7,SC8} atomic Fermi
gases with population imbalance loaded in optical
lattices\cite{SF7,SF8,FG1,FG2,FG3,FG4,FG5,FG6,FG7,
FG8,FG9,FG10,FG11,FG12,FG13,FG14,FG15,FG16,FG17,FG18} and harmonic
traps, \cite{HTa} and dense quark matter,\cite{DMa} shows that the
FF and LO states compete with a number of other states, such as the Sarma
($\textbf{q}=0$) states,\cite{Sa} and the superfluid-normal separation phase (also known as the phase separation phase).\cite{FD1,FD2,FD3,FD4}
It turns out that in some regions of
momentum space the FF (or LO) phase provides the minimum of the mean-field
expression of the Helmholtz free energy. The phase diagrams have been calculated for atomic Fermi gases in free space \cite{FDFS1,FDFS2,FDFS3}; it was found that the parameter window for FF (or LO) states is extremely narrow.
In contrast,  a considerable parameter window for the existence of the FF phase has been found for population-imbalanced (but not mass-imbalanced) mixtures in an optical lattice. \cite{SF9,SF10,SF12,SF18} Phase diagrams for a $^6Li-$ $^{40}K$ mixture  at zero temperature were obtained in  Ref. [\onlinecite{FDFS4}], but the calculations were limited to the emergence
of insulating phases during the evolution of superfluidity
from the BCS to the BEC regime, and the  competition between the FF and Sarma phases was
ignored. The polarization versus temperature diagram,\cite{SZ} presented in Fig. 2,
shows that there are three phases: the Sarma phase, the FF phase, and the normal phase in which the Helmholtz free energy is
minimized for gapless phase. The zero polarization line is the conventional Bardeen-Cooper-Schrieffer state. Contrary to the phase diagram of population-imbalanced
$^6Li$ Fermi gas, where the phase separation appears for low polarizations,
it was found\cite{SZ} the existence of a polarization window for the FF phase. This means that
as soon as the system is polarized it goes into the FF phase if the temperature is low
enough. This polarization window is larger for a majority of $^{40}K$ atoms compared to
the majority of $^6Li$ atoms.

In what follows, we calculate the collective mode dispersion of the FF states numerically  using the system parameters corresponding to a point in the above-mentioned polarization window for the FF phase:
$f_\uparrow=0.225$, $f_\downarrow=0.275$, $U/J_\uparrow=2$ and $T/J_\uparrow=0.01$. The FF vector $\textbf{q}=(q_x,q_y)$, the gap and the chemical potentials in the mean-field approximation are defined by the solution of the mean-field equations (see Eqs. (\ref{SPGFSL3}) in Sec. III): $\textbf{q}=(0.0489\pi/a,0)$, $\Delta_\textbf{q}/J_\uparrow=0.3668 $,  $\mu_\uparrow/J_\uparrow=2.0909$, and $\mu_\downarrow/J_\uparrow=0.5561$. For the Sarma state the corresponding mean-field values are as follows: $\Delta/J_\uparrow=0.3635 $,  $\mu_\uparrow/J_\uparrow=2.0466$, and $\mu_\downarrow/J_\uparrow=0.5918$.  The FF phase is the most stable as it provides the minimum of the mean-field expression of the Helmholtz free energy (the ratio between the FF free energy and the Sarma free energy is about 0.9986).

The paper is organized as follows. In the next Section we apply the functional-integral formalism to derive equations for the single-particle excitations and for the two-particle collective modes. In Section III, we numerically solve the BS equation to obtain the  Nambu-Goldstone  excitation spectrum and the corresponding speed of sound of  an  interacting Fermi mixture of $^6Li-$ $^{40}K$ atoms in a two-dimensional optical lattice at finite
temperatures  with the Fulde-Ferrell order parameter. In the last Section we discuss the difference between the  collective-mode dispersion, obtained by means of $4\times 4$ secular determinant, and obtained by the $16\times 16$ secular determinant.

\section{ BETHE-SALPETER APPROACH TO THE COLLECTIVE MODES}
\subsection{The functional-integral technique}

  The Green's functions in the  functional-integral approach are defined by means
  of the so-called generating functional with sources for the boson and fermion  fields.
  In our problem,  the corresponding functional integrals  cannot be evaluated
exactly because the interaction part of the Hamiltonian (\ref{H}) is
quartic in the Grassmann fermion fields. However, we can transform
the quartic terms to a quadratic form by introducing  a model system
which consists of a four-component boson field $A_{\alpha}(z)$ ($\alpha=1,2,3,4,\quad  z=(\textbf{r}_i,v),\quad 0\leq v \leq \beta)$
 interacting with fermion fields
$\widehat{\overline{\psi}} (y)=\widehat{\Psi}^\dag (y)/\sqrt{2}$ and
$\widehat{\psi}(x)=\widehat{\Psi}(x)/\sqrt{2}$.  The action of this model system is assumed to
be of the following form
 $S= S^{(F)}_0+S^{(B)}_0+S^{(F-B)}$, where:
$$S^{(F)}_0=\widehat{\overline{\psi}
}(y)\widehat{G}^{(0)-1}(y;x)\widehat{\psi} (x),$$
$$S^{(B)}_0=\frac{1}{2}A_{\alpha}(z)D^{(0)-1}_{\alpha
\beta}(z,z')A_{\beta}(z'),$$$$ S^{(F-B)}=\widehat{\overline{\psi}}
(y)\widehat{\Gamma}^{(0)}_{\alpha}(y,x\mid z)\widehat{\psi}
(x)A_{\alpha}(z).$$

 The action $S^{(F)}_0$ describes the fermion  part of the system.
 The generalized inverse Green's function
 of free fermions
$\widehat{G}^{(0)-1}(y;x)$ is given by the following  diagonal matrix:
\begin{equation}\begin{split}
&\widehat{G}^{(0)-1}(y;x)=\\&\sum_{\textbf{k},\omega_m}\exp\left[\imath
\textbf{k.}(\textbf{r}_i-\textbf{r}_{i'})-\omega_m(u-u')\right]
G_{n_1n_2}^{(0)-1}(\textbf{k},\imath\omega_m) , \nonumber
\end{split}\end{equation}
 where  $G_{11}^{(0)-1}(\textbf{k},\imath\omega_m)=
 -G_{33}^{(0)-1}(-\textbf{k},-\imath\omega_m)
 $, and
$-G_{22}^{(0)-1}(-\textbf{k},-\imath\omega_m)=G_{44}^{(0)-1}(\textbf{k},\imath\omega_m)$.
The symbol $\sum_{\omega_m}$ is used to denote $\beta^{-1}\sum_{m}$
(for fermion fields $\omega_m=
   (2\pi/\beta)(m +1/2) ;m=0, \pm 1, \pm 2,... $). In the case of the FF states of a population-imbalanced Fermi gas,  the non-interacting Green's
   function is:
   \begin{widetext}
   \begin{equation}
 \widehat{G}^{(0)-1}(\textbf{k},\imath\omega_m)=\left(%
\begin{array}{cccc}
  \imath\omega_m-\xi_\uparrow(\textbf{k})&0&0&0  \\
  0&\imath\omega_m-\xi_\downarrow(\textbf{k}) &0&0  \\
 0 &0 &\imath\omega_m+\xi_\uparrow(\textbf{k})&0\\
 0 &0 & 0&\imath\omega_m+\xi_\downarrow(\textbf{k})
\end{array}%
\right),
 \nonumber\end{equation}
 where $\xi_{\uparrow,\downarrow}(\textbf{k})=2J_{\uparrow,\downarrow}(1-\cos k_x)+2J_{\uparrow,\downarrow}(1-\cos k_y)-\mu_{\uparrow,\downarrow}$.

 The action $S^{(B)}_0$ describes
the boson field which mediates
the  fermion-fermion on-site interaction in the Hubbard Hamiltonian. The bare boson propagator in $S^{(B)}_0$  is defined as:
\begin{equation}\widehat{D}^{(0)}
(z,z')=\delta(v-v')U\delta{j,j'}\left(%
\begin{array}{cccc}
  0&1&0&0  \\
 1 &0 &0&0  \\
 0 &0 &0&0\\
 0 &0 & 0&0
\end{array}%
\right).\nonumber\end{equation}  The Fourier transform of the boson
propagator is given by
\begin{equation}
\widehat{D}^{(0)} (z,z')=\frac{1}{N} \sum_\textbf{k}\sum_{\omega_p}
 e^{\left\{\imath\left[\textbf{k.}\left(\textbf{r}_j-\textbf{r}_{j'}\right)
 -\omega_p\left(v-v'\right)\right]\right\}}\widehat{D}^{(0)}(\textbf{k}),
\widehat{D}^{(0)}(\textbf{k})= \left(%
\begin{array}{cccc}
  0&U &0&0  \\
 U  &0 &0&0  \\
 0 &0 &0&0 \\
 0 &0 & 0&0
\end{array}%
\right).\label{FTD0}\end{equation}

 The interaction between the fermion and the boson
 fields is described by the action $S^{(F-B)}$.
 The bare vertex
$\widehat{\Gamma}^{(0)}_{\alpha}(y_1;x_2\mid
z)=\widehat{\Gamma}^{(0)}_{\alpha}(i_1,u_1;i_2, u_2\mid
j,v)=\delta(u_1-u_2)\delta(u_1-v)\delta_{i_1i_2}\delta_{i_1j}\widehat{\Gamma}^{(0)}(\alpha)$
is a $4\times 4$ matrix, where
\begin{equation}
\widehat{\Gamma}^{(0)}(\alpha)=\frac{1}{2}(\gamma_0+\alpha_z)\delta_{\alpha1}
+\frac{1}{2}(\gamma_0-\alpha_z)\delta_{\alpha2}+
\frac{1}{2}(\alpha_x+\imath\alpha_y)\delta_{\alpha3}+
\frac{1}{2}(\alpha_x-\imath\alpha_y)\delta_{\alpha4}.\label{Gamma0}\end{equation}
The Dirac matrix $\gamma_0$  and the  matrices
 $\widehat{\alpha}_i$  are defined as (when a four-dimensional space is used,\cite{KM} the electron spin operators $\sigma_i$ has to be replaced by $\widehat{\alpha}_i  \gamma_0$):
$$\gamma_0=\left(%
\begin{array}{cccc}
  1&0&0&0  \\
 0&1&0&0  \\
 0& 0& -1&0  \\
 0& 0& 0&-1  \\
\end{array}%
\right),\quad  \widehat{\alpha}_i=\left(%
\begin{array}{cc}
  \sigma_i & 0  \\
 0& \sigma_y\sigma_i\sigma_y \\
\end{array}%
\right), i=x,y,z.$$

The relation between the Hubbard  model and our model system can be
demonstrated by applying the Hubbard-Stratonovich transformation for
the fermion operators:
\begin{equation}\int \mu[A]\exp\left[\widehat{\overline{\psi}}
(y)\widehat{\Gamma}^{(0)}_{\alpha}(y;x|z)\widehat{\psi}(x)A_{\alpha}(z)\right]
=\exp\left[-\frac{1}{2}\widehat{\overline{\psi}}
(y)\widehat{\Gamma}^{(0)}_{\alpha}(y;x|z)\widehat{\psi}(x)
D_{\alpha,\beta}^{(0)}(z,z') \widehat{\overline{\psi}}
(y')\widehat{\Gamma}^{(0)}_{\beta}(y';x'|z')\widehat{\psi}(x')\right].\label{HSa}\end{equation}
 The functional measure $D\mu[A]$ is chosen to be:
$$
\mu[A]=DAe^{-\frac{1}{2}A_{\alpha}(z)D_{\alpha,\beta}^{(0)-1}(z,z')
A_{\beta}(z')},\int \mu[A] =1.$$

 According to the field-theoretical approach, the expectation value of a general operator
$\widehat{O}(u)$ can be expressed as a functional integral over the
boson field $A$ and the Grassmann fermion fields
$\widehat{\overline{\psi}}$ and $\widehat{\psi}$:
\begin{equation}<\widehat{T}_u(\widehat{O}(u))>=\frac{1}{Z[J,M]}\int
D\mu[\widehat{\overline{\psi}},\widehat{\psi},A]\widehat{O}(u)
\exp\left[J_{\alpha}(z)A_{\alpha}(z)-M(\widehat{\overline{\psi}},\widehat{\psi})\right]|_{J=M=0},\label{Ex}\end{equation}
where the symbol $<...>$ means that the thermodynamic average is
made. The
functional $Z[J,M]$ is defined by
\begin{equation}
Z[J,M]=\int
D\mu[\widehat{\overline{\psi}},\widehat{\psi},A]
\exp\left[J_{\alpha}(z)A_{\alpha}(z)-M(\widehat{\overline{\psi}},\widehat{\psi})\right],\label{GW}
\end{equation}where the functional measure
$D\mu[\widehat{\overline{\psi}},\widehat{\psi},A]=DAD\widehat{\overline{\psi}}D\widehat{\psi}
\exp\left(S\right)$ satisfies the condition $\int
D\mu[\widehat{\overline{\psi}},\widehat{\psi},A]=1$. The quantity
$J_\alpha(z)$ is the source of the boson field, while the sources
$M_{ij}(y;x)$ of the fermion fields are included in the
$M(\widehat{\overline{\psi}},\widehat{\psi})$ term :
\begin{eqnarray} &M(\widehat{\overline{\psi}},\widehat{\psi})=
\psi^\dag_\uparrow(y)
M_{11}(y;x)\psi_\uparrow(x)+\psi^\dag_\downarrow(y)
M_{21}(y;x)\psi_\uparrow(x)+\psi^\dag_\uparrow(y) M_{12}(y;
x)\psi_\downarrow(x)+\psi^\dag_\downarrow(y)
M_{22}(y;x)\psi_\downarrow(x)\nonumber\\& + \psi_\uparrow(y)
M_{31}(y;x)\psi_\uparrow(x)+\psi_\downarrow(y)
M_{41}(y;x)\psi_\uparrow(x)+\psi_\uparrow(y)
M_{32}(y;x)\psi_\downarrow(x)+\psi_\downarrow(y)
M_{42}(y;x)\psi_\downarrow(x)\nonumber\\& + \psi^\dag_\uparrow(y)
M_{13}(y;x) x)\psi^\dag_\uparrow(x)+\psi^\dag_\downarrow(y)
M_{23}(y;x)\psi^\dag_\uparrow(x)+\psi^\dag_\uparrow(y)
M_{14}(y;x)\psi^\dag_\downarrow(x)+\psi^\dag_\downarrow(y)
M_{24}(y;x)\psi^\dag_\downarrow(x)\nonumber\\& +\psi_\uparrow(y)
M_{33}(y;x)\psi^\dag_\uparrow(x)+\psi_\downarrow(y)
M_{43}(y;x)\psi^\dag_\uparrow(x)+\psi_\uparrow(y)
M_{34}(y;x)\psi^\dag_\downarrow(x)+\psi_\downarrow(y)
M_{44}(y;x)\psi^\dag_\downarrow(x).
 \label{M1}\end{eqnarray}
Here, we have introduced complex indices $1=\{n_1,x_1\}$, and
$2=\{n_2,y_2\}$.

We
shall now use  a functional derivative $\delta / \delta M(2;1)$;
depending on the spin degrees of freedom, there are sixteen possible
derivatives.  By means of the definition (\ref{Ex}), one can express all Green's functions in terms of the functional derivatives with respect to the
  corresponding  sources
 of the
generating functional of the connected Green's functions $W[J,M]=\ln
Z[J,M]$. Thus, we define the following Green's and vertex functions
which will be used to  analyze the collective modes of our model:

  The Boson Green's
function is $D_{\alpha \beta}(z,z')$ is a $4\times 4$ matrix defined
as $D_{\alpha \beta}(z,z')=-\frac{\delta^2W}{\delta
J_{\alpha}(z)\delta J_{\beta}(z')}$.

The generalized single-fermion Green's function $G_{n_1n_2}(x_1;y_2)$ is the
 matrix (\ref{EGF}) whose elements are
$G_{n_1n_2}(x_1;y_2)=-\delta W/\delta M_{n_2n_1}(y_2;x_1)$.
 Depending on the two spin degrees of freedom,
$\uparrow$ and $\downarrow$, there exist eight "normal" Green's
functions  and eight "anomalous" Green's functions. We introduce
Fourier transforms of the "normal"
$G_{\sigma_1,\sigma_2}(\textbf{k},u_1-u_2)=
-<\widehat{T}_u(\psi_{\sigma_1,\textbf{k}}(u_1)\psi^\dag_{\sigma_2,\textbf{k}}(u_2))>$,
and "anomalous"
$F_{\sigma_1,\sigma_2}(\textbf{k},u_1-u_2)=-<\widehat{T}_u(\psi_{\sigma_1,\textbf{k}}(u_1)
\psi_{\sigma_2,-\textbf{k}}(u_2))>$ one-particle Green's functions,
where $\{\sigma_1,\sigma_2\}=\uparrow,\downarrow$. Here
$\psi^+_{\uparrow,\textbf{k}}(u),\psi_{\uparrow,\textbf{k}}(u)$ and
$\psi^+_{\downarrow,\textbf{k}}(u),\psi_{\downarrow,\textbf{k}}(u)$
are the creation-annihilation Heisenberg operators. The Fourier
transform of the generalized single-particle Green's function is given by
\begin{equation}
\widehat{G}(1;2)=
\frac{1}{N}\sum_{\textbf{k}}\sum_{\omega_{m}}\exp\{\imath\left[\textbf{k.}\left(
\textbf{r}_{i_1}-\textbf{r}_{i_2}\right)-\omega_{m}(u_1-u_2)\right]\}
\left(%
\begin{array}{cc}
  \widehat{G}(\textbf{k},\imath\omega_m) & \widehat{F}(\textbf{k},\imath\omega_m)  \\
 \widehat{F}^\dag(\textbf{k},\imath\omega_m) & -\widehat{G}(-\textbf{k},-\imath\omega_m) \\
\end{array}%
\right).\label{FTGF}\end{equation} Here, $\widehat{G}$ and
$\widehat{F}$ are $2\times 2$ matrices whose elements are
$G_{\sigma_1,\sigma_2}$ and $F_{\sigma_1,\sigma_2}$, respectively.

The two-particle
Green's function $K\left(%
\begin{array}{cc}
  n_1,x_1 & n_3,y_3  \\
  n_2,y_2 & n_4,x_4 \\
\end{array}%
\right)$ is defined as
\begin{equation}
K\left(%
\begin{array}{cc}
  n_1,x_1 & n_3,y_3  \\
  n_2,y_2 & n_4,x_4 \\
\end{array}%
\right)=K\left(%
\begin{array}{cc}
  1 & 3  \\
  2 & 4 \\
\end{array}%
\right)=\frac{\delta^2 W}{\delta M_{n_2n_1}(y_2;x_1)\delta
M_{n_3n_4}(y_3;x_4)}=-\frac{\delta G_{n_1n_2}(x_1;y_2)}{\delta
M_{n_3n_4}(y_3;x_4)} . \label{TGF1}
\end{equation}
This definition of $K$ allows us to conclude that if the
approximation used for $G$ is chosen in accordance with the recipes
proposed by Baym and Kadanoff,\cite{BK2} then $K$ is automatically
conserving.

 The vertex function $\widehat{\Gamma}_{\alpha}(2;1 \mid
  z)$ for a given $\alpha$  is a $4 \times 4$ matrix whose elements are:
\begin{equation}
\widehat{\Gamma}_{\alpha}(i_2,u_2;i_1,u_1 \mid
v,j)_{n_2n_1}=-\frac{\delta G_{n_2n_1}^{-1}(i_2,u_2;i_1,u_1)}{\delta
J_{\beta}(z')} D^{-1}_{\beta \alpha}(z',z). \label{VF}\end{equation}
\subsection{Equations of the boson and fermion Green's functions }
It is well-known that the fermion self-energy (fermion mass
operator) $\widehat{\Sigma}(1;2)$ can be defined by means of the
so-called SD equations. They can be derived using the fact that the
measure $D\mu[\overline{\psi},\psi,A]$ is invariant under the
translations $\overline{\psi}\rightarrow
\overline{\psi}+\delta\overline{\psi}$ and  $A\rightarrow A+\delta
A$:
\begin{equation}
D^{(0)-1}_{\alpha
\beta}(z,z')R_\beta(z')+\frac{1}{2}Tr\left(\widehat{G}(1;2)\widehat{\Gamma}^{(0)}_{\alpha}(2;1\mid
z)\right)+J_\alpha(z)=0, \label{SE1}
\end{equation}
\begin{equation}
\widehat{G}^{-1}(1;2)-\widehat{G}^{(0)-1}(1;2)+\widehat{\Sigma}(1;2)+\widehat{M}(1;2)=0,
\label{SE2}
\end{equation}
where $R_\alpha(z)=\delta W/\delta J_\alpha(z)$ is the average boson
field. The fermion self-energy   $\widehat{\Sigma}$, is a $4\times
4$ matrix which can be written as a sum of Hartree
$\widehat{\Sigma}^H$ and Fock $\widehat{\Sigma}^F$ parts. The
Hartree part is a diagonal matrix whose elements are:
\begin{equation}
\Sigma^H(i_1,u_1;i_2,u_2)_{n_1n_2}=\frac{1}{2}
\widehat{\Gamma}_{\alpha}^{(0)}(i_1,u_1;i_2,u_2|j,v)_{n_1n_2}
D^{(0)}_{\alpha\beta}(j,v;j',v')\widehat{\Gamma}_{\beta}^{(0)}(i_3,u_3;i_4,u_4|j',v')_{n_3n_4}
G_{n_4n_3}(i_4,u_4;i_3,u_3) .\label{H1}
\end{equation}

The Fock part of the fermion self-energy is given by:
\begin{equation}\begin{split}&
\Sigma^F(i_1,u_1;i_2,u_2)_{n_1n_2}=-
\widehat{\Gamma}_{\alpha}^{(0)}(i_1,u_1;i_6,u_6|j,v)_{n_1n_6}
D^{(0)}_{\alpha\beta}(j,v;j',v')\widehat{\Gamma}_{\beta}^{(0)}(i_4,u_4;i_5,u_5|j',v')_{n_4n_5}
\times\\&K\left(%
\begin{array}{cc}
  n_5,i_5,u_5 & n_3,i_3,u_3  \\
  n_4,i_4,u_4 & n_6,i_6,u_6 \\
\end{array}%
\right)G^{-1}_{n_3n_2}(i_3,u_3;i_2,u_2).\label{Fock
sigma}\end{split}
\end{equation}
The Fock part of the fermion self-energy depends on the two-particle
Green's function $K$; therefore the SD equations and the BS equation
for $K$ have to be solved self-consistently.

Our approach to the Hubbard model allows us to obtain exact
equations of the Green's functions by using the field-theoretical
technique. We now wish to return to our statement that the Green's
functions are the thermodynamic average of the
$\widehat{T}_u$-ordered products of field operators. The standard
procedure for calculating the Green's functions, is to apply Wick's
theorem. This  enables us to evaluate the $\widehat{T}_u$-ordered
products of field operators as a perturbation expansion involving
only wholly contracted field operators. These expansions can be
summed formally to yield different equations of Green's functions.
The main disadvantage of this procedure is that the validity of the
equations must be verified diagram by diagram. For this reason we
will  use the method of Legendre transforms of the generating
functional for connected Green's functions.\cite{DM} By applying the
same steps as  in Ref. [\onlinecite{ZKexc}] we obtain the BS
equation of the two-particle Green's function, the Dyson equation of
the boson Green's function, and the vertex equation:
\begin{equation}
 K^{-1}\left(%
\begin{array}{cc}
  n_2,i_2,u_2 & n_3,i_3,u_3  \\
  n_1,i_1,u_1 & n_4,i_4,u_4 \\
\end{array}%
\right)= K^{(0)-1}\left(%
\begin{array}{cc}
  n_2,i_2,u_2 & n_3,i_3,u_3  \\
  n_1,i_1,u_1 & n_4,i_4,u_4 \\
\end{array}%
\right)-I\left(%
\begin{array}{cc}
  n_2,i_2,u_2 & n_3,i_3,u_3  \\
  n_1,i_1,u_1 & n_4,i_4,u_4 \\
\end{array}%
\right),\label{BSK}
\end{equation}
\begin{equation}
D_{\alpha \beta}(z,z')=D^{(0)}_{\alpha \beta}(z,z')+D^{(0)}_{\alpha
\gamma}(z,z'')\Pi_{\gamma\delta}(z'',z''')D^{(0)}_{\delta
\beta}(z,z'),\label{BosonDyson}\end{equation}
\begin{equation}\begin{split}
&\widehat{\Gamma}_{\alpha}(i_2,u_2;i_1,u_1\mid
z)_{n_2n_1}=\widehat{\Gamma}^{(0)}_{\alpha}(i_2,u_2;i_1,u_1\mid
z)_{n_2n_1}+I\left(%
\begin{array}{cc}
 n_2,i_2,u_2 & n_3,i_3,u_3  \\
  n_1,i_1,u_1 & n_4,i_4,u_4 \\
\end{array}%
\right)\times\\&
K^{(0)}\left(%
\begin{array}{cc}
 n_3,i_3,u_3 & n_6,i_6,u_6  \\
  n_4,i_4,u_4 & n_5,i_5,u_5 \\
\end{array}%
\right)\widehat{\Gamma}_{\alpha}(i_6,u_6;i_5,u_5\mid z)_{n_6n_5}.
\label{Edward}\end{split}\end{equation}
Here, $$K^{(0)}\left(%
\begin{array}{cc}
 n_2,i_2,u_2 & n_3,i_3,u_3  \\
  n_1,i_1,u_1 & n_4,i_4,u_4 \\
\end{array}%
\right)=G_{n_2n_3}(i_2,u_3;i_2,u_2)G_{n_4n_1}(i_4,u_4;i_1,u_1)$$  is
the two-particle free propagator constructed from a pair of fully
dressed generalized single-particle Green's functions. The kernel
$I=\delta\Sigma/\delta G$ of the BS equation can be expressed as a
functional derivative  of the fermion self-energy
$\widehat{\Sigma}$. Since
$\widehat{\Sigma}=\widehat{\Sigma}^H+\widehat{\Sigma}^F$, the BS
kernel $I=I_{exc}+I_d$ is a sum of
 functional derivatives of the Hartree $\Sigma^H$ and
Fock $\Sigma^F$ contributions to the self-energy:
\begin{equation}
I_{exc}\left(%
\begin{array}{cc}
 n_2,i_2,u_2 & n_3,i_3,u_3  \\
  n_1,i_1,u_1 & n_4,i_4,u_4 \\
\end{array}%
\right)=\frac{\delta\Sigma^H(i_2,u_2;i_1,u_1)_{n_2n_1}}{\delta
G_{n_3n_4}(i_3,u_3;i_4,u_4)},\quad
I_d\left(%
\begin{array}{cc}
 n_2,i_2,u_2 & n_3,i_3,u_3  \\
  n_1,i_1,u_1 & n_4,i_4,u_4 \\
\end{array}%
\right)=\frac{\delta\Sigma^F(i_2,u_2;i_1,u_1)_{n_2n_1}}{\delta
G_{n_3n_4}(i_3,u_3;i_4,u_4)}.\label{Kernel}\end{equation} The
general response function $\Pi$ in the Dyson equation
(\ref{BosonDyson}) is defined as
\begin{equation}\Pi_{\alpha \beta}(z,z')=
\widehat{\Gamma}^{(0)}_{\alpha}(i_1,u_1;i_2,u_2 \mid
z)_{n_1n_2}K\left(%
\begin{array}{cc}
  n_2,i_2,u_2 & n_3,i_3,u_3  \\
  n_1,i_1,u_1 & n_4,i_4,u_4 \\
\end{array}%
\right)\widehat{\Gamma}^{(0)}_{\beta}(i_3,u_3,i_4,u_4\mid
z')_{n_3n_4} .\label{Pi}
\end{equation}
 The  functions $D$, $K$  and $\widehat{\Gamma}$
 are related by the  identity:
\begin{equation}\begin{split}&
K^{(0)}\left(%
\begin{array}{cc}
 n_2,i_2,u_2 & n_3,i_3,u_3  \\
  n_1,i_1,u_1 & n_4,i_4,u_4 \\
\end{array}%
\right)\widehat{\Gamma}_{\beta}(i_4,u_4;i_3,u_3\mid
z')_{n_4n_3}D_{\beta \alpha}(z',z)\\&= K\left(%
\begin{array}{cc}
  n_2,i_2,u_2 & n_3,i_3,u_3  \\
  n_1,i_1,u_1 & n_4,i_4,u_4 \\
\end{array}%
\right)\widehat{\Gamma}^{(0)}_{\beta}(i_4,u_4;i_3,u_3\mid
z')_{n_4n_3}D^{(0)}_{\beta \alpha}(z',z), \label{GammaK}
\end{split}\end{equation}

 By introducing the boson proper self-energy $P^{-1}_{\alpha
\beta}(z,z')=
\Pi^{-1}_{\alpha\beta}(z,z')+D^{(0)}_{\alpha\beta}(z,z')$ one can
rewrite the Dyson equation (\ref{BosonDyson})
 for the boson Green's function as:
\begin{equation}D^{-1}_{\alpha \beta}(z,z') =D^{(0)-1}_{\alpha
\beta}(z,z')-P_{\alpha \beta}(z,z').\label{DE}\end{equation} The
proper self-energy and the vertex function $\widehat{\Gamma}$ are
related by the following equation:
 \begin{equation}\begin{split}&
P_{\alpha
\beta}(z,z')=\frac{1}{2}Tr\left[\widehat{\Gamma}_\alpha^{(0)}(y_1,x_2|z)\widehat{G}(x_2,y_3)
\widehat{\Gamma}_\beta(y_3,x_4|z')\widehat{G}(x_4,y_1)\right]\\&=\frac{1}{2}
\widehat{\Gamma}^{(0)}_{\alpha}(i_1,u_1;i_2,u_2\mid
z)_{n_1n_2}G_{n_2n_3}(i_2,u_2;i_3,u_3)
\widehat{\Gamma}_{\beta}(i_3,u_3;i_4,u_4\mid
z')_{n_3n_4}G_{n_4n_1}(i_4,u_4;i_1,u_1).
\label{PG}\end{split}\end{equation} It is also possible to express
the proper self-energy in terms of the two-particle Green's function
$\widetilde{K}$ which satisfies the BS equation
$\widetilde{K}^{-1}=K^{(0)-1}-I_d$, but its kernel
$I_d=\delta\Sigma^F/\delta G$ includes only diagrams that represent
the direct interactions:
  \begin{equation}\begin{split}&
P_{\alpha
\beta}(z,z')=\widehat{\Gamma}^{(0)}_{\alpha}(i_1,u_1;i_2,u_2\mid
z)_{n_1n_2}\widetilde{K}\left(%
\begin{array}{cc}
  n_2,i_2,u_2 & n_3,i_3,u_3  \\
  n_1,i_1,u_1 & n_4,i_4,u_4 \\
\end{array}%
\right) \widehat{\Gamma}^{(0)}_{\beta}(i_3,u_3;i_4,u_4\mid
z')_{n_3n_4}\\&
=\widehat{\Gamma}^{(0)}_{n_1n_2}(\alpha)\widetilde{K}\left(%
\begin{array}{cc}
  n_2,\textbf{r}_j,v & n_3,\textbf{r}_{j'},v'  \\
  n_1,\textbf{r}_j,v & n_4,\textbf{r}_{j'},v' \\
\end{array}\right)\widehat{\Gamma}^{(0)}_{n_3n_4}(\beta). \label{P}\end{split}\end{equation}
One can obtain the spectrum of the collective excitations as poles of the boson Green's function by solving the Dyson equation (\ref{DE}), but one has first to deal with the BS equation for the function $\widetilde{K}$. In other words, this method involves two steps. For this reason, it is easy to  obtain the collective modes by locating the poles of the two-particle Green's function $K$ using the solutions of the corresponding BS equation.

 \subsection{Mean-field approximation for the generalized single-particle Green's functions}
  As we have already mentioned, the BS equation and
the SD equations  have to be solved self-consistently. In what follows, we use an approximation which allows us to decouple the
above-mentioned equations and to obtain a linearized integral
equation for the Fock term. To apply  this approximation we first
use Eq. (\ref{GammaK}) to rewrite the Fock term as
\begin{equation}\Sigma^F(i_1,u_1;i_2,u_2)_{n_1n_2}=-
\widehat{\Gamma}_{\alpha}^{(0)}(i_1,u_1;i_3,u_3|j,v)_{n_1n_3}
D_{\alpha\beta}(j,v;j',v')G_{n_3n_4}(i_3,u_3;i_4,u_4)\widehat{\Gamma}_{\beta}(i_4,u_4;i_2,u_2|j',v')_{n_4n_2}
,\label{MassFock}\end{equation} and after that we replace $D$ and
$\widehat{\Gamma}$ in (\ref{MassFock}) by the free boson propagator
$D^{(0)}$  and by the bare vertex $\widehat{\Gamma}^{(0)}$,
respectively.  In this approximation the Fock term assumes the form:
\begin{equation}\begin{split}
&\Sigma_0^F(i_1,u_1;i_2,u_2)_{n_1n_2}=-
\widehat{\Gamma}_{\alpha}^{(0)}(i_1,u_1;i_3,u_3|j,v)_{n_1n_3}
D^{(0)}_{\alpha\beta}(j,v;j',v')\widehat{\Gamma}_{\beta}^{(0)}(i_4,u_4;i_2,u_2|j',v'))_{n_4n_2}
G_{n_3n_4}(i_3,u_3;i_4,u_4)=\\&
-U\delta_{i_1,i_2}\delta(u_1-u_2)\left(%
\begin{array}{cccc}
  0 &G_{12}(1;2)&0&-G_{14}(1;2)  \\
 G_{21}(1;2)& 0 &-G_{23}(1;2)&0  \\
 0 & -G_{32}(1;2)&0&G_{34}(1;2)  \\
 -G_{41}(1;2) & 0&G_{43}(1;2)&0  \\
\end{array}%
\right).\label{SF1}
\end{split}\end{equation}
The total self-energy is $\widehat{\Sigma}(i_1,u_1;i_2,u_2)=\widehat{\Sigma}^H(i_1,u_1;i_2,u_2)+\widehat{\Sigma}^F(i_1,u_1;i_2,u_2)$, where
\begin{equation}\begin{split}&
\widehat{\Sigma}^H(i_1,u_1;i_2,u_2)=\\&\frac{U}{2}\delta_{i_1,i_2}\delta(u_1-u_2)\left(%
\begin{array}{cccc}
  G_{22}(1;2)-G_{44}(1;2) &0&0&0 \\
 0& G_{11}(1;2)-G_{33}(1;2) &0&0  \\
 0 & 0&G_{44}(1;2)-G_{22}(1;2)&0  \\
 0& 0&0)&G_{33}(1;2)-G_{11}(1;2)  \\
\end{array}%
\right)\end{split} \label{HFG}\end{equation} \end{widetext}The
contributions to $\Sigma(i_1,u_1;i_2,u_2)$, due to the elements on
the major diagonal of the above matrices, will be included into the
chemical potential. To obtain an analytical
expression for the generalized single-particle Green's function, we  assume
two more approximations. First, since the experimentally relevant magnetic fields are not strong enough to cause
spin flips, we shall neglect
$G_{12}=G_{21}=G_{34}=G_{43}=0$. Second, we neglect the
frequency dependence of the Fourier transform of the Fock part of
the fermion self-energy. Thus, the Dyson equation for the
generalized single-particle Green's function becomes:
\begin{widetext}
$$\widehat{G}^{-1}(1;2)=\left(%
\begin{array}{cccc}
  G^{(0)-1}_{11}(1;2) &0&0&-\Delta e^{\imath2\textbf{q.r}_{i_1}}\delta(\textbf{r}_{i_1}-\textbf{r}_{i_2}) \\
 0& G^{(0)-1}_{22}(1;2) &\Delta e^{\imath2\textbf{q.r}_{i_1}}\delta(\textbf{r}_{i_1}-\textbf{r}_{i_2})&0  \\
 0 & \Delta e^{-\imath2\textbf{q.r}_{i_1}}\delta(\textbf{r}_{i_1}-\textbf{r}_{i_2}) &G^{(0)-1}_{33}(1;2)&0  \\
 -\Delta e^{-\imath2\textbf{q.r}_{i_1}}\delta(\textbf{r}_{i_1}-\textbf{r}_{i_2}) & 0&0&G^{(0)-1}_{44}(1;2)  \\
\end{array}%
\right)$$ \end{widetext} We can eliminate the phase factors  $ e^{\imath2\textbf{q.r}_{i_1}}$
by performing the unitary transformation between the old generalized
single-particle Green's function $\widehat{G}$ and the new one
$\widehat{\widetilde{G}}$, i.e.
$\widehat{\widetilde{G}}(1;2)=U(1).\widehat{G}(1;2).U^\dag(2)$, where  the
corresponding matrix is as follows
$$ U(1)=\left(%
\begin{array}{cccc}
  e^{-\imath\textbf{q.r}_{i_1}} &0&0&0\\
 0& -e^{-\imath\textbf{q.r}_{i_1}} &0&0  \\
 0 & 0 &-e^{\imath\textbf{q.r}_{i_1}}&0  \\
 0& 0&0&e^{\imath\textbf{q.r}_{i_1}}  \\
\end{array}%
\right).$$ After performing this unitary transformation, the Green's
function $\widehat{\widetilde{G}}(1;2)$  become functions of
$\widehat{\widetilde{G}}(\textbf{r}_{i_1}-\textbf{r}_{i_2};u_1-u_2)$,
and the corresponding Fourier transform is:
\begin{equation}\begin{split}&
\widehat{\widetilde{G}}(1;2)=\sum_{\textbf{k},\omega_m}\exp\left[\imath(\textbf{k.}
(\textbf{r}_{i_1}-\textbf{r}_{i_2})-\omega_m(u_1-u_2)\right]
\times\\&\left(%
\begin{array}{cccc}
  \widetilde{G}_{11}(\textbf{k},\imath\omega_m) &0&0&\widetilde{G}_{14}(\textbf{k},\imath\omega_m) \\
 0& \widetilde{G}_{22}(\textbf{k},\imath\omega_m)&\widetilde{G}_{23}(\textbf{k},\imath\omega_m)&0  \\
 0 & \widetilde{G}_{32}(\textbf{k},\imath\omega_m)&\widetilde{G}_{33}(\textbf{k},\imath\omega_m)&0  \\
 \widetilde{G}_{41}(\textbf{k},\imath\omega_m)& 0&0&\widetilde{G}_{44}(\textbf{k},\imath\omega_m)  \\
\end{array}%
\right) \label{GF3}\end{split}\end{equation}
\section{ Collective modes of  $^6Li-$ $^{40}K$ mixture }
 The FF superfluid state is expected to occur on
the BCS side of the Feshbach resonance, where the effective
attractive interaction between fermion atoms leads to BCS type
pairing.
In the case when  the order parameter is assumed to vary as a single plane wave,
we have a broken translational
invariance, and as a result, the normal and anomalous Green's
functions have phase factors associated with the FF quasimomentum $\textbf{q}$, which can be eliminated using the previously mentioned unitary transformation.

In the mean filed approximation, the FF vector $\textbf{q}$, as well as the chemical potentials $\mu_\uparrow$ and $\mu_\downarrow$, and the gap $\Delta_{\textbf{q}}$ are defined by the solutions of following set of four equations (the number equations, the gap equation  and the q-equation):\cite{SF18}
\begin{widetext}
\begin{equation}\begin{split}&
f_\uparrow=\frac{1}{N}\sum_\textbf{k}
\left[u_\textbf{q}^2(\textbf{k})f(\omega_+(\textbf{k},\textbf{q}))+
v_\textbf{q}^2(\textbf{k})f(-\omega_-(\textbf{k},\textbf{q}))\right],
\quad f_\downarrow=\frac{1}{N}\sum_\textbf{k}
\left[u_\textbf{q}^2(\textbf{k})f(\omega_-(\textbf{k},\textbf{q}))+
v_\textbf{q}^2(\textbf{k})f(-\omega_+(\textbf{k},\textbf{q}))\right],
\\& 1=\frac{U}{N}\sum_\textbf{k}
\frac{1-f(\omega_-(\textbf{k},\textbf{q}))-
f(\omega_+(\textbf{k},\textbf{q}))}{2E_\textbf{q}(\textbf{k})},
\\& 0=\frac{1}{N}\sum_\textbf{k}\left\{ \frac{\partial
\eta_\textbf{q}(\textbf{k})}{\partial
q_x}\left[f(\omega_+(\textbf{k},\textbf{q}))-f(\omega_-(\textbf{k},\textbf{q}))\right]
+ \frac{\partial \chi_\textbf{q}(\textbf{k})}{\partial
q_x}\left[1-\frac{\chi_\textbf{q}(\textbf{k})}{E_\textbf{q}(\textbf{k})}
\left[1-f(\omega_+(\textbf{k},\textbf{q}))-f(\omega_-(\textbf{k},\textbf{q}))\right]
\right]\right\},\\&
 \chi_{\textbf{q}}(\textbf{k})=\frac{1}{2}\left[\xi_\uparrow(\textbf{q}+\textbf{k})+
\xi_\downarrow(\textbf{q}-\textbf{k})\right],\quad
\eta_{\textbf{q}}(\textbf{k})=\frac{1}{2}\left[\xi_\uparrow(\textbf{q}+\textbf{k})-
\xi_\downarrow(\textbf{q}-\textbf{k})\right],\quad
\omega_\pm(\textbf{k},\textbf{q})= E_{\textbf{q}}(\textbf{k})\pm
\eta_{\textbf{q}}(\textbf{k}),
\\&u_{\textbf{q}}^2(\textbf{k})=
\frac{1}{2}\left[1+\frac{\chi_{\textbf{q}}(\textbf{k})}{E_{\textbf{q}}(\textbf{k})}\right],
v_{\textbf{q}}^2(\textbf{k})=
\frac{1}{2}\left[1-\frac{\chi_{\textbf{q}}(\textbf{k})}{E_{\textbf{q}}(\textbf{k})}\right],
\quad
E_{\textbf{q}}(\textbf{k})=\sqrt{\chi_{\textbf{q}}^2(\textbf{k})+
\Delta_{\textbf{q}}^2},\label{SPGFSL3}\end{split}\end{equation}
The Fourier transform of the generalized single-particle Green's function in the mean field approximation is as follows:
\begin{equation}\begin{split}&
\widehat{\widetilde{G}}^{\textbf{q}}(\textbf{k},\imath\omega_m)=\left(%
\begin{array}{cccc}
  \frac{\imath\omega_m+\xi_\downarrow(\textbf{q}-\textbf{k})}{A(\textbf{q},\textbf{k},\imath\omega_m)} &0&0&\frac{\Delta_{\textbf{q}}}{A(\textbf{q},\textbf{k},\imath\omega_m)} \\
 0& \frac{\imath\omega_m+\xi_\uparrow(\textbf{q}-\textbf{k})}{B(\textbf{q},\textbf{k},\imath\omega_m)}&
 - \frac{\Delta_{\textbf{q}}}{B(\textbf{q},\textbf{k},\imath\omega_m)}&0  \\
 0 & -\frac{\Delta_{\textbf{q}}}{B(\textbf{q},\textbf{k},\imath\omega_m)}&
 \frac{\imath\omega_m-\xi_\downarrow(\textbf{q}+\textbf{k})}{B(\textbf{q},\textbf{k},\imath\omega_m)}&0  \\
 \frac{\Delta_{\textbf{q}}}{A(\textbf{q},\textbf{k},\imath\omega_m)} & 0&0&\frac{\imath\omega_m-\xi_\uparrow(\textbf{q}+\textbf{k})}{A(\textbf{q},\textbf{k},\imath\omega_m)}  \\
\end{array}%
\right),\\&
A(\textbf{q},\textbf{k},\imath\omega_m)=\left(\imath\omega_m-\xi_\uparrow(\textbf{q}+\textbf{k})\right)
\left( \imath\omega_m+\xi_\downarrow(\textbf{q}-\textbf{k})\right)-\Delta_{\textbf{q}}^2,\\&
B(\textbf{q},\textbf{k},\imath\omega_m)=\left(\imath\omega_m-\xi_\downarrow(\textbf{q}+\textbf{k})\right)
\left( \imath\omega_m+\xi_\uparrow(\textbf{q}-\textbf{k})\right)-\Delta_{\textbf{q}}^2.
\label{GFFFa}\end{split}\end{equation}
\end{widetext}
The spectrum of the collective modes will be obtained by solving the BS equation  in the GRPA. As we
have already mentioned, the kernel of the BS equation is a sum of
the direct $I_d=\delta \Sigma^F /\delta G$ and exchange
$I_{exc}=\delta \Sigma^H /\delta G$ interactions, written as
derivatives of the Fock (\ref{SF1}) and the Hartree (\ref{HFG}) parts of the self-energy. Thus, in the GRPA  the corresponding equation for the
BS amplitude $\Psi^{\textbf{Q}}_{n_2,n_1}=\int
\frac{d\Omega}{2\pi}\int
\frac{d^d\textbf{k}}{(2\pi)^d}\Psi^{\textbf{Q}}_{n_2,n_1}(\textbf{k};\Omega)$
can be obtained from Eq. (\ref{BSEdZ}) by  performing  integration
over the momentum vectors:
\begin{widetext}
\begin{equation}\Psi^{\textbf{Q}}_{n_2n_1}=K^{(0)}\left(%
\begin{array}{cc}
  n_1 & n_3  \\
  n_2 & n_4 \\
\end{array}%
|\omega(\textbf{Q})\right)
\left[I_d\left(%
\begin{array}{cc}
  n_3 & n_5  \\
  n_4 & n_6 \\
\end{array}%
\right)+I_{exc}\left(%
\begin{array}{cc}
  n_3 & n_5  \\
  n_4 & n_6 \\
\end{array}%
\right)\right]\Psi^{\textbf{Q}}_{n_6,n_5},\label{BSEdZ1}
\end{equation}
where the two-particle propagator $K^{(0)}$ and  the direct and
exchange interactions are defined  as follows:
\begin{equation}\begin{split}&K^{(0)}\left(%
\begin{array}{cc}
  n_1 & n_3  \\
  n_2 & n_4 \\
\end{array}%
|\omega(\textbf{Q})\right)\equiv K_{n_1n_3n_4n_2}=\int
\frac{d\Omega}{2\pi}
 \int\frac{d^d\textbf{k}}{(2\pi)^d}\widetilde{G}_{n_1n_3}
 \left(\textbf{k}+\textbf{Q},\Omega+\omega(\textbf{Q})\right)\widetilde{G}_{n_4n_2}(\textbf{k},
 \Omega),\\
&I_d\left(%
\begin{array}{cc}
  n_1 & n_3  \\
  n_2 & n_4 \\
\end{array}%
\right)=-\Gamma^{(0)}_\alpha(n_1,n_3)D^{(0)}_{\alpha\beta}
\Gamma^{(0)}_\beta(n_4,n_2),
\\&I_{exc}\left(%
\begin{array}{cc}
  n_1 & n_3  \\
  n_2 & n_4 \\
\end{array}%
\right)=
\frac{1}{2}\Gamma^{(0)}_\alpha(n_1,n_2)D^{(0)}_{\alpha\beta}
\Gamma^{(0)}_\beta(n_4,n_3).\label{BSEdZ2}
\end{split}\end{equation}

The BS equation (\ref{BSEdZ1}) written in the matrix form is
$\left(\widehat{I}+U\widehat{Z}\right)\widehat{\Psi}=0$, where
 $\widehat{I}$ is the unit matrix,  the matrix $\widehat{Z}$ is a $16 \times 16$ matrix, and
 the transposed matrix of
$\widehat{\Psi}$  is given by:
$$\widehat{\Psi}^T=
\left(%
\begin{array}{cccccccccccccccc}
  \Psi^{\textbf{Q}}_{1,1} &
  \Psi^{\textbf{Q}}_{1,2} &
  \Psi^{\textbf{Q}}_{1,3} &
  \Psi^{\textbf{Q}}_{1,4} &
  \Psi^{\textbf{Q}}_{2,1} &
  \Psi^{\textbf{Q}}_{2,2} &
  \Psi^{\textbf{Q}}_{2,3} &
  \Psi^{\textbf{Q}}_{2,4} &
  \Psi^{\textbf{Q}}_{3,1} &
  \Psi^{\textbf{Q}}_{3,2} &
  \Psi^{\textbf{Q}}_{3,3} &
  \Psi^{\textbf{Q}}_{3,4} &
  \Psi^{\textbf{Q}}_{4,1} &
  \Psi^{\textbf{Q}}_{4,2} &
  \Psi^{\textbf{Q}}_{4,3} &
  \Psi^{\textbf{Q}}_{4,4} \\
\end{array}%
\right).
$$
\end{widetext}
The $16 \times 16$  secular determinant $det|\widehat{I}+U\widehat{Z}|$ can be
rewritten as a block diagonal determinant:
\begin{equation}
\left|%
\begin{array}{cccc}
\widehat{D}_{8\times 8}&0&0\\
0&\widehat{D}_{4\times 4}&0\\
0&0&\widehat{1}_{4\times 4}\\
\end{array}%
\right|,\label{BD}\end{equation} where  $\widehat{1}_{4\times 4}$ is
a $4\times 4$ unit matrix. The block structure of the secular
determinant allows us to separate the sixteen BS amplitudes into
three independent groups related to the blocks $\widehat{D}_{8\times
8}$, $\widehat{D}_{4\times 4}$, and   $\widehat{1}_{4\times 4}$. The
determinant
\begin{widetext}
\begin{equation}
Z_8=\left|%
\begin{array}{cccccccc}
U^{-1} -\frac{K_{1414}}{2}&K_{1411}&\frac{K_{1111}}{2}&0&0&
 \frac{ K_{1414}}{2}&K_{1114}&-\frac{K_{1111}}{2}\\
   -\frac{K_{4414}}{2}&U^{-1}+K_{4411}&\frac{K_{1411}}{2}&0&0&
 \frac{K_{4414}}{2}&K_{1414}&-\frac{K_{1411}}{2}\\
  \frac{H_{4444}}{2}&0&U^{-1}-\frac{H_{1414}}{2}&H_{1444}&
 H_{4414}&-\frac{H_{4444}}{2}&0& \frac{H_{1414}}{2}\\
 \frac{H_{1444}}{2}&0&-\frac{H_{1114}}{2}&U^{-1}+H_{1144}&
 H_{1414}&-\frac{H_{1444}}{2}&0&
 \frac{H_{1114}}{2}\\
  \frac{H_{4414}}{2}&0&-\frac{K_{1411}}{2}&H_{1414}&U^{-1}+H_{4411}
  &-\frac{H_{4414}}{2}&0
  &\frac{H_{1411}}{2}\\
 \frac{H_{1414}}{2}&0&-\frac{H_{1111}}{2}&H_{1114}&H_{1411}&U^{-1}-\frac{H_{1414}}{2}&0
 &\frac{H_{1111}}{2}\\
  -\frac{K_{1444}}{2}&K_{1414}&\frac{K_{1114}}{2}&0&0&\frac{K_{1444}}{2}&U^{-1}+K_{1144}
  &-\frac{K_{1114}}{2}\\
 -\frac{K_{4444}}{2}&K_{4414}&\frac{K_{1414}}{2}&0&0&\frac{K_{4444}}{2}&K_{1444}
 &U^{-1}-\frac{K_{1414}}{2}
 \\
 \\
 \end{array}%
\right| \label{FF88}\end{equation}
of the  $\widehat{D}_{8\times 8}$ block determines the amplitudes $
\Psi^{\textbf{Q}}_{1,1}, \Psi^{\textbf{Q}}_{1,4},
\Psi^{\textbf{Q}}_{2,2}, \Psi^{\textbf{Q}}_{2,3},
\Psi^{\textbf{Q}}_{3,2}, \Psi^{\textbf{Q}}_{3,3},
\Psi^{\textbf{Q}}_{4,1}$ and $\Psi^{\textbf{Q}}_{4,4}$.  The other four amplitudes
$ \Psi^{\textbf{Q}}_{1,2}, \Psi^{\textbf{Q}}_{2,1},
\Psi^{\textbf{Q}}_{3,4}$ and $ \Psi^{\textbf{Q}}_{4,3}$ are related to the  $\widehat{D}_{4\times 4}$ block.
 The last four amplitudes are equal
to zero, i.e.  $ \Psi^{\textbf{Q}}_{1,3}=
\Psi^{\textbf{Q}}_{2,4}=\Psi^{\textbf{Q}}_{3,1}=\Psi^{\textbf{Q}}_{4,2}=0$. The
collective-mode dispersion is defined by the secular determinant
(\ref{FF88}).
 At  a finite temperature the
elements of $Z_8$  are:
\begin{equation}\begin{split}
&K_{1111}=\frac{1}{2}\left(I_{m,m}+I_{\widetilde{\gamma},\widetilde{\gamma}}
-2J_{\widetilde{\gamma},m}-L_{l,l}-L_{\gamma,\gamma}+2K_{\gamma,l}\right),
K_{4444}=\frac{1}{2}\left(I_{m,m}+I_{\widetilde{\gamma},\widetilde{\gamma}}+
2J_{\widetilde{\gamma},m}-L_{l,l}-L_{\gamma,\gamma}-2K_{\gamma,l}\right),\\&
K_{1144}=\frac{1}{2}\left(I_{l,l}+I_{\gamma,\gamma}+2J_{\gamma,l}
-L_{m,m}-L_{\widetilde{\gamma},\widetilde{\gamma}}-2K_{\widetilde{\gamma},m}\right),
K_{4411}=\frac{1}{2}\left(I_{l,l}+I_{\gamma,\gamma}-2J_{\gamma,l}
-L_{m,m}-L_{\widetilde{\gamma},\widetilde{\gamma}}+2K_{\widetilde{\gamma},m}\right),\\&
K_{1114}=\frac{1}{2}\left(-I_{l,m}+I_{\gamma,\widetilde{\gamma}}+
J_{l,\widetilde{\gamma}}-J_{\gamma,m}-L_{l,m}+L_{\gamma,\widetilde{\gamma}}+
K_{\gamma,m}-K_{l,\widetilde{\gamma}}\right),\\&
K_{4414}=\frac{1}{2}\left(I_{l,m}-I_{\gamma,\widetilde{\gamma}}+
J_{l,\widetilde{\gamma}}-J_{\gamma,m}
+L_{l,m}-L_{\gamma,\widetilde{\gamma}}+
K_{\gamma,m}-K_{l,\widetilde{\gamma}}\right)\\&
K_{1414}=\frac{1}{2}\left(I_{l,l}-I_{\gamma,\gamma}-
L_{m,m}+L_{\widetilde{\gamma},\widetilde{\gamma}}\right),
K_{1444}=\frac{1}{2}\left(I_{l,m}+I_{\gamma,\widetilde{\gamma}}+
J_{l,\widetilde{\gamma}}+J_{\gamma,m}+L_{l,m}+L_{\gamma,\widetilde{\gamma}}
+K_{\gamma,m}+K_{l,\widetilde{\gamma}}\right),\\&
K_{1411}=\frac{1}{2}\left(-I_{l,m}-I_{\gamma,\widetilde{\gamma}}+
J_{l,\widetilde{\gamma}}+J_{\gamma,m}
-L_{l,m}-L_{\gamma,\widetilde{\gamma}}
+K_{\gamma,m}+K_{l,\widetilde{\gamma}}\right),
\nonumber\end{split}\end{equation}
where the symbols $I,J,L$ and $K$ are defined  as:\cite{SF18}
\begin{equation}\begin{split}&
I_{a,b}=\frac{1}{2N}\sum_\textbf{k}a^\textbf{q}_{\textbf{k},\textbf{Q}}b^\textbf{q}_{\textbf{k},\textbf{Q}}
\left[\frac{1-f\left(\omega_-(\textbf{k},\textbf{q})
\right)-f\left(\omega_+(\textbf{k}+\textbf{Q},\textbf{q})
\right)}{\omega+\Omega_\textbf{q}(\textbf{k},\textbf{Q})-\varepsilon_\textbf{q}(\textbf{k},\textbf{Q})]}
-\frac{1-f\left(\omega_+(\textbf{k},\textbf{q})
\right)-f\left(\omega_-(\textbf{k}+\textbf{Q},\textbf{q})
\right)}{\omega+\Omega_\textbf{q}(\textbf{k},\textbf{Q})+\varepsilon_\textbf{q}(\textbf{k},\textbf{Q})]}\right]
,\\&
J_{a,b}=\frac{1}{2N}\sum_\textbf{k}a^\textbf{q}_{\textbf{k},\textbf{Q}}b^\textbf{q}_{\textbf{k},\textbf{Q}}
\left[\frac{1-f\left(\omega_-(\textbf{k},\textbf{q})
\right)-f\left(\omega_+(\textbf{k}+\textbf{Q},\textbf{q})
\right)}{\omega+\Omega_\textbf{q}(\textbf{k},\textbf{Q})-\varepsilon_\textbf{q}(\textbf{k},\textbf{Q})]}
+\frac{1-f\left(\omega_+(\textbf{k},\textbf{q})
\right)-f\left(\omega_-(\textbf{k}+\textbf{Q},\textbf{q})
\right)}{\omega+\Omega_\textbf{q}(\textbf{k},\textbf{Q})+\varepsilon_\textbf{q}(\textbf{k},\textbf{Q})]}\right]
,\\&
K_{a,b}=\frac{1}{2N}\sum_\textbf{k}a^\textbf{q}_{\textbf{k},\textbf{Q}}b^\textbf{q}_{\textbf{k},\textbf{Q}}
\left[\frac{f\left(\omega_-(\textbf{k},\textbf{q})
\right)-f\left(\omega_-(\textbf{k}+\textbf{Q},\textbf{q})
\right)}{\omega+\Omega_\textbf{q}(\textbf{k},\textbf{Q})
+\epsilon_\textbf{q}(\textbf{k},\textbf{Q})]}
+\frac{f\left(\omega_+(\textbf{k},\textbf{q})
\right)-f\left(\omega_+(\textbf{k}+\textbf{Q},\textbf{q})
\right)}{\omega+\Omega_\textbf{q}(\textbf{k},\textbf{Q})
-\epsilon_\textbf{q}(\textbf{k},\textbf{Q})]}\right]
,\\&
L_{a,b}=\frac{1}{2N}\sum_\textbf{k}a^\textbf{q}_{\textbf{k},
\textbf{Q}}b^\textbf{q}_{\textbf{k},\textbf{Q}}
\left[\frac{f\left(\omega_-(\textbf{k},\textbf{q})
\right)-f\left(\omega_-(\textbf{k}+\textbf{Q},\textbf{q})
\right)}{\omega+\Omega_\textbf{q}(\textbf{k},\textbf{Q})+
\epsilon_\textbf{q}(\textbf{k},\textbf{Q})]}
-\frac{f\left(\omega_+(\textbf{k},\textbf{q})
\right)-f\left(\omega_+(\textbf{k}+\textbf{Q},\textbf{q})
\right)}{\omega+\Omega_\textbf{q}(\textbf{k},\textbf{Q})
-\epsilon_\textbf{q}(\textbf{k},\textbf{Q})]}\right]
.\label{IJKL}\end{split}\end{equation} \end{widetext}
Here,
$\varepsilon_\textbf{q}(\textbf{k},\textbf{Q})=
 E_\textbf{q}(\textbf{k}+\textbf{Q})+ E_\textbf{q}(\textbf{k})$,
  $\epsilon_\textbf{q}(\textbf{k},\textbf{Q})=
 E_\textbf{q}(\textbf{k}+\textbf{Q})- E_\textbf{q}(\textbf{k})$,  $\Omega_\textbf{q}(\textbf{k},\textbf{Q})=\eta_\textbf{q}(\textbf{k})-\eta_\textbf{q}
 (\textbf{k}+\textbf{Q})$,
 and $a$ and $b$ are one of the following form factors:
\begin{equation}\begin{split}&\gamma^\textbf{q}_{\textbf{k},\textbf{Q}}=
u^\textbf{q}_{\textbf{k}}u^\textbf{q}_{\textbf{k}+\textbf{Q}}
+v^\textbf{q}_{\textbf{k}}v^\textbf{q}_{\textbf{k}+\textbf{Q}},
l^\textbf{q}_{\textbf{k},\textbf{Q}}=u^\textbf{q}_{\textbf{k}}
u^\textbf{q}_{\textbf{k}+\textbf{Q}}
-v^\textbf{q}_{\textbf{k}}v^\textbf{q}_{\textbf{k}+\textbf{Q}},\nonumber\\&
\widetilde{\gamma}^\textbf{q}_{\textbf{k},\textbf{Q}}=
u^\textbf{q}_{\textbf{k}}v^\textbf{q}_{\textbf{k}+\textbf{Q}}
-u^\textbf{q}_{\textbf{k}+\textbf{Q}}v^\textbf{q}_{\textbf{k}},
 m^\textbf{q}_{\textbf{k},\textbf{Q}}=
u^\textbf{q}_{\textbf{k}}v^\textbf{q}_{\textbf{k}+\textbf{Q}}+
u^\textbf{q}_{\textbf{k}+\textbf{Q}}v^\textbf{q}_{\textbf{k}}.\nonumber\end{split}\end{equation}
The elements $H$ are defined by
$H_{ijkl}(\textbf{q},\textbf{Q},\omega)=K_{ijkl}(\textbf{q},-\textbf{Q},-\omega)$.
The secular determinant $det|\widehat{D}_{8\times 8}|$ also provides the gap equation in the limit
$\textbf{Q}\rightarrow 0$ and $\omega\rightarrow 0$. Thus, our Hubbard-Stratonovich transformation is in
accordance with the canonical mean-field approximation.
\begin{figure}\includegraphics[scale=0.7]{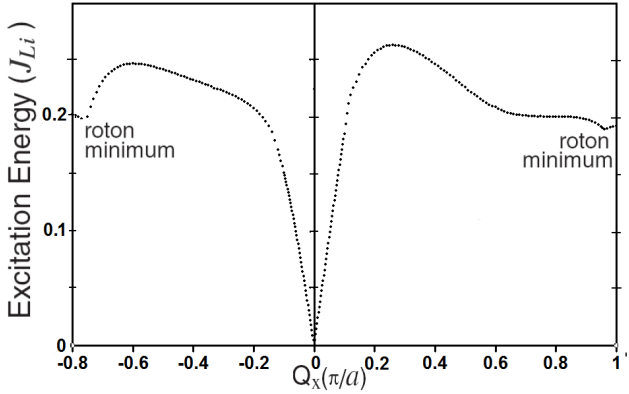}
    \caption{The collective-mode dispersion of $^6Li-$ $^{40}K$ mixture in 2D optical lattice along
the positive $(Q_x,0)$ and negative $(-Q_x, 0)$ directions. The
filling factors are $f_{Li} = 0.225$ and $f_K= 0.275$. The interaction
strength and the temperature are $U/J_{Li} = 2$ and
$T/J_{Li}= 0.01$, respectively.
The secular determinant $Z_8$ provides the speed of
sound to the positive direction $c = 0.614 J_{Li}a/\hbar$, while the
speeds of sound to the negative direction is $c = 0.534 J_{Li}a/\hbar$
.}\end{figure}

In Fig. 3, we have presented the dispersion relation $\omega(\textbf{Q})$  calculated for the system parameters listed in Sec. I.  The FF vector $\textbf{Q}$ is directed along the x
axis. The speed of sound, $c$, to the positive and negative directions of the
$Q_x$  axis is defined by  $d\omega(Q_x)/dQ_x$ at $Q_x\rightarrow 0$. For the dispersions presented in Fig. 3, we obtain $c = 0.80 J_{Li}a/\hbar$ in positive direction, and   $c = 0.54 J_{Li}a/\hbar$ in the  negative direction. There are two roton minima at $Q_x=0.96 \pi/a$ and at $Q_x=-0.76 \pi/a$.
 \begin{figure} \includegraphics[scale=0.7]{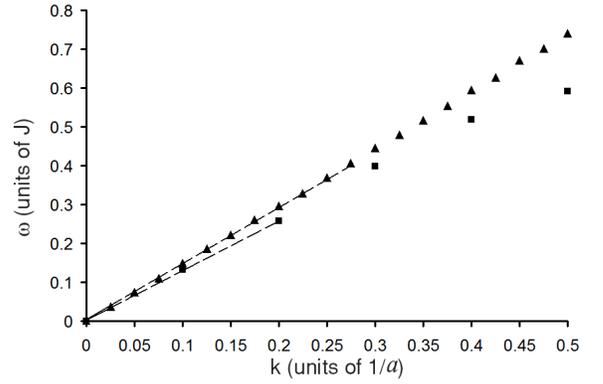}
    \caption{The collective-mode dispersion  $\omega(\textbf{k})$ (triangles) calculated by the Bethe-Salpeter formalism for a 2D system with $\textbf{q}=q_x\textbf{e}_x$, $q_x=0.06 \pi/a$, $\mu_\uparrow=3.5 J$, $\mu_\downarrow=2.5 J$, $\Delta=0.27 J$, $U=3J$ and  $T=0.07J$.  The speed of sound to the direction of the x axis is $c=1.48 Ja/\hbar$. The squire points are taken from Ref. [\onlinecite{SF19}], and the corresponding speed of sound is  $c=1.29 Ja/\hbar$.   }\end{figure}
\section{Discussion}
We have  studied the  Nambu-Goldstone  excitation spectrum and the corresponding speed of sound of  an  interacting Fermi mixture of Lithium-6 and
Potassium-40 atoms Fermi gases in deep
optical lattices by using the Bethe-Salpeter equation in the GRPA. The generalized single-particle Green's function, used in our numerical calculations, takes into account  all possible   thermodynamic averages. In view of the fact that most of the previous numerical calculations are based on the Nambu-Gor'kov single-particle Green's function (\ref{G2}) which leads to the $4\times 4$ secular determinant, one may well ask whether the collective-mode dispersion, defined by the secular determinant $Z_4$, is  significantly different in comparison to  the dispersion obtained with the secular determinant $Z_8$.

To answer this question, we have calculated the collective-mode dispersion relation $\omega(Q_x)$  using  the $4\times 4$ determinant (\ref{SecDet4}). The numerical results show  that the  determinant  $Z_4$ and the  determinant  $Z_8$, both  provide almost the same dispersions; the difference is about $2\%-7\%$ in the interval $-0.1\pi/a<Q_x<0.1\pi/a$, and less than $1\%$  out of this interval.

In conclusion, we briefly discuss  the collective-mode dispersion of a population-imbalanced atomic Fermi gas obtained in Ref. [\onlinecite{SF19}] by applying the Kadanoff-Baym density response formalism. To the best of our knowledge this is the first paper in which the collective-mode dispersion and the corresponding speed of sound are calculated starting from the generalized single-particle Green's function. However, it is not difficult to prove analytically that the location of the  poles of the  density response function is provided by the secular determinant $Z_4$. Since the Fock contributions to the electron self-energy are the same in the BS and the Kadanoff-Baym approaches, the only reason to have $Z_4$ instead of $Z_8$ is in the corresponding Hartree contributions.  According to the  Ref. [\onlinecite{SF19}], the Hartree contributions are $-UG_{22}(1;2)$, $-UG_{11}(1;2))$, $-UG_{44}(1;2)$, $-UG_{33}(1;2)$, while the Hartree terms  $(U/2)(G_{44}(1;2)-G_{22}(1;2))$, $(U/2)(G_{33}(1;2)-G_{11}(1;2))$, $(U/2)(G_{22}(1;2)-G_{44}(1;2))$, and $(U/2)(G_{11}(1;2)-G_{33}(1;2))$ are obtained from the SD equations without any approximation. In other words, the Hartree contribution to the electron self-energy, as defined by Eq. (\ref{H1}), is an exact result.  The  dispersion obtained by means of the secular determinant $Z_8$ is presented in Fig. 4. In the range of small $k$, we find a difference of  about $15\%$  between the speed of sound obtained by means of $Z_8$ secular determinant, and the speed of sound, calculated in Ref. [\onlinecite{SF19}]. The figure also indicates that the difference between the two dispersion curves tends to increase with $k$, reaching $25\%$ at $k=0.5/a$.


\begin{thebibliography}{999}
%
\bibitem{FF} P. Fulde, and R. A. Ferrell, Phys. Rev. \textbf{135}, A550 (1964).
%
\bibitem{LO} A. I. Larkin, and Y. N.
Ovchinnikov,  Zh. Eksp. Teor. Fiz., \textbf{47}, 1136 (1964) [Sov.
Phys. JETP \textbf{20}, 762 (1965)].
%
\bibitem{N} Y. Nambu, Phys. Rev. Lett., \textbf{4},380 (1960).
%
\bibitem{G} J. Goldstone, Nuovo Cimento, \textbf{19},154 (1961).
%
\bibitem{SFexp1} J. K. Chin, D. E. Miller, Y. Liu, C. Stan, W. Setiawan, C.
Sanner, K. Xu, and W. Ketterle, Nature (London) \textbf{443}, 961
(2006).
%
\bibitem{SFexp2} M.W. Zwierlein, A. Schirotzek, C. H. Schunck, and  W. Ketterle,
Science \textbf{311}, 492 (2006).
%
\bibitem{SFexp3} G. B. Partridge, W. Li, R. I. Kamar, Y.
Liao, and R. G. Hulet, Science \textbf{311}, 503 (2006).
%
\bibitem{SFexp4} M.W. Zwierlein, C. H. Schunck, A. Schirotzek, and W. Ketterle, Nature
(London) \textbf{442}, 54 (2006).
%
\bibitem{SFexp5} Y. Shin, C. Schunck, A. Schirotzek, and
W. Ketterle, Nature (London) \textbf{451}, 689 (2008).
%
\bibitem{SFexp6} Y.
Liao, A. S. C. Rittner, T. Paprotta, W. Li, G. B. Partridge, R. G.
Hulet, S. K. Baur, and E. J. Mueller, Nature (London) \textbf{467},
567 (2010).
%
\bibitem{SFexp7} S. Nascimb`ene, N. Navon, K. J. Jiang, L. Tarruell, M.
Teichmann, J. McKeever, F. Chevy, and C. Salomon, Phys. Rev. Lett.
\textbf{103}, 170402 (2009).
%
\bibitem{SFexp8} S. Nascimb`ene, N. Navon, K. Jiang, F.
Chevy, and C. Salomon, Nature (London) \textbf{463}, 1057 (2010).
%
\bibitem{SF0} G. M. Bruun and B.R. Mottelson, Phys. Rev. Lett. \textbf{87}, 270403 (2001).
%
\bibitem{SF1} W. Hofstetter, J. I. Cirac, P. Zoller, E. Demler, and M. D.
Lukin, Phys. Rev. Lett. \textbf{89}, 220407 (2002).
%
\bibitem{SF1a} Y. Ohashi,  and A. Griffin, Phys. Rev. A \textbf{67}, 063612 (2003).
%
\bibitem{SF2} G.
Orso and G.V. Shlyapnikov, Phys. Rev. Lett. \textbf{95}, 260402
(2005).
%
\bibitem{SF3} L. P. Pitaevskii, S. Stringari, and G. Orso,
Phys. Rev. A \textbf{71}, 053602 (2005).
%
\bibitem{SF4} W. Hofstetter Phil. Mag.,
\textbf{86}, 1891 (2006).
%
\bibitem{SF5} W. Yi and L.-M. Duan, Phys. Rev. A
\textbf{73}, 063607 (2006).
%
\bibitem{SF6} T. Koponen, J.-P. Martikainen, J. Kinnunen, and P.
T\"{o}rm\"{a}1, Phys. Rev. A \textbf{73}, 033620 (2006).
%
\bibitem{SF7} T. Koponen, J.-P. Martikainen, J. Kinnunen, L> M. Jensen, and P.
T\"{o}rm\"{a}1, New Journal of Physics \textbf{8}, 179 (2006).
%
\bibitem{SF8} M. Iskin and C. A. R. S\'{a} de Melo,
Phys. Rev. Lett. \textbf{99}, 080403 (2007).
%
\bibitem{SF9} T. K. Koponen, T.
Paananen, J.-P. Martikainen, and P. T\"{o}rm\"{a}, Phys. Rev. Lett.
\textbf{99}, 120403 (2007).
%
\bibitem{SF9b} M. M. Parish, S. K. Baur, E. J. Mueller, and D. A. Huse, Phys. Rev.
Lett. \textbf{99}, 250403 (2007).
%
\bibitem{SF9a} R. Haussmann, W. Rantner, S. Cerrito, and W. Zwerger, Phys. Rev.
A \textbf{75}, 023610 (2007).
%
 \bibitem{SF10} T. Paananen, T. K. Koponen,
P. T\"{o}rm\"{a}, and J. P. Martikainen, Phys. Rev. A \textbf{77}, 053602
(2008).
%
\bibitem{SF11} I. Bloch, J. Dalibard, and W. Zwerg, Rev. Mod. Phys.
\textbf{80}, 885 (2008).
%
\bibitem{SF12} T. Paananen, J. Phys. B: At. Mol. Opt.
Phys. \textbf{42}, 165304 (2009).
%
\bibitem{SF13}  Ai-Xia Zhang and Ju-Kui Xue,
Phys. Rev. A \textbf{80}, 043617 (2009).
%
\bibitem{SF14}   T. K. Koponen, T.
Paananen, and P. T\"{o}rm\"{a}l, Phys. Rev. Lett. \textbf{102},
165301 (2009).
%
\bibitem{SF15} J. M. Edge and N. R. Cooper, Phys. Rev.
Lett. \textbf{103}, 065301 (2009).
%
\bibitem{SF15a} M. Iskin and C. A. R. S\'{a} de Melo, Phys. Rev.
Lett. \textbf{103}, 165301 (2009).
%
\bibitem{SF16} Y. Yunomae, I. Danshita, D.
Yamamoto, N Yokoshi, and S Tsuchiya, Journal of Physics: Conference
Series \textbf{150}, 032128 (2009).
%
\bibitem{SF16a} Y. Yunomae,  D.
Yamamoto, I. Danshita, N Yokoshi, and S Tsuchiya,  Phys. Rev. A
\textbf{80}, 063627 (2009).
%
\bibitem{SF17} R. Ganesh, A. Paramekanti and A. A. Burkov, Phys. Rev. A
\textbf{80}, 043612 (2009).
%
\bibitem{SF17a} Cheng Zhao, Lei Jiang, Xunxu Liu, W. M. Liu, Xubo Zou, and Han Pu, Phys. Rev. A
\textbf{81}, 063642 (2010).
%
\bibitem{SF18} Z. Koinov, R. Mendoza and M. Fortes,  Phys. Rev. Lett. \textbf{106}
100402 (2011).
%
\bibitem{SF19} M. O. J. Heikkinen and P. T\"{o}rm\"{a},  Phys. Rev. A \textbf{83},
053630 (2011).
%
\bibitem{SF19a} L. M. Sieberer and M. A. Baranov,  Phys. Rev. A \textbf{84},
063633 (2011).
%
\bibitem{SF19b} J. Kajala, F. Massel, and P. T\"{o}rm\"{a},  Phys. Rev. A \textbf{84},
041601(R)(2011).
%
\bibitem{SF19c} Dong-Hee Kim, and P. T\"{o}rm\"{a},  Phys. Rev. B \textbf{85},
180508(R)(2012).
%
\bibitem{SF19d} M. O. J. Heikkinen, Dong-Hee Kim, and P. T\"{o}rm\"{a},  Phys. Rev. B \textbf{87},
224513 (2013).
%
\bibitem{SF19e} M. Iskin, Phys. Rev.
A \textbf{88}, 013631 (2013).
%
\bibitem{SF19g} K. Seo, C. Zhang, and S. Tewari, Phys. Rev.
A \textbf{88}, 063601 (2013).
%
\bibitem{SF20} R. Mendoza, M. Fortes, M. A. Sol\'{\i}s and Z. Koinov,   Phys. Rev. A. \textbf{88}
033606 (2013).
%
\bibitem{SF21} A. Korolyuk, J. J. Kinnunen, and P. T\"{o}rm\"{a},  Phys. Rev. B \textbf{89},
013602  (2014).
%
\bibitem{SF22} Shaoyu Yin,1 J.-P. Martikainen, and P. T\"{o}rm\"{a},  Phys. Rev. B \textbf{89},
014507  (2014).
%
\bibitem{Schwinger} J. Schwinger, Phys. Rev., \textbf{82}, 914
(1951).
%
\bibitem{Dyson} F. J. Dyson, Phys. Rev., \textbf{75}, 1736
(1949).
%
\bibitem{BetheS}  H. A. Bethe and E. E. Salpeter, Phys. Rev., \textbf{82}, 309
(1951); ibit.  \textbf{84}, 1232 (1951).
%
%
\bibitem{CGexc} R. Cot\^{e} and A. Griffin, Phys. Rev. B \textbf{37}, 4539
(1988).
%
\bibitem{CG1} R. Cot\^{e} and A. Griffin, Phys. Rev. B, \textbf{48}, 10404 (1993).
%
 \bibitem{CCexc} H. Chu and Y. C. Chang, Phys. Rev. B \textbf{54}, 5020
(1996).
%
\bibitem{ZKexc} Z. Koinov, Phys. Rev. B \textbf{72}, 085203 (2005).
%
\bibitem{Com} R. Combescot, M. Yu. Kagan, and S. Stringari, Phys. Rev. A
\textbf{74}, 042717 (2006).
%
\bibitem{ZGK} Z. G. Koinov,  Physica Status Solidi (B), \textbf{247} ,140
(2010); Physica C \textbf{407}, 470 (2010).
%
\bibitem{ZK1} Z. G. Koinov, Ann. Phys. (Berlin) \textbf{522}, 693 (2010).
%
\bibitem{PA} P. W. Anderson, Phys. Rev. \textbf{112}, 1900 (1958).
%
 \bibitem{R} G. Rickayzen, Phys. Rev. \textbf{115}, 795 (1959).
 %
\bibitem{BR} L. Belkhir and M. Randeria, Phys. Rev. B \textbf{49}, 6829
(1994).
%
\bibitem{BK} G. Baym and L. P. Kadanoff, Phys. Rev., \textbf{124}, 287 (1961).
%
 \bibitem{BK2} G. Baym, Phys. Rev. \textbf{127}, 1391 (1962).
%
\bibitem{ZR} C. Mac\^{e}do, and M. Coutinho-Filho, Phys. Rev. B, \textbf{43}, 13515 (1991).
%
\bibitem{DM} C. De Dominicis and P. Martin, J. Math. Phys., \textbf{5}, 430 (1964).
%
\bibitem{IZ} C. Itzykson and J. Zuber, Quantum Field Theory, McGraw-Hill, NY 1980.
%
\bibitem{KM} K. Maki, p. 1035, in "Superconductivity", edited by R.D. Parks, Marcel Dekker, Inc., New York, (1969).
%
\bibitem{SC1} P. Pieri, D. Neilson, and G. C. Strinati, Phys. Rev. B
\textbf{75}, 113301 (2007).
%
\bibitem{SC2} T. Hakio\u{g}lu and M. \c{S}ahin, Phys.
Rev. Lett. \textbf{98}, 166405 (2007).
%
\bibitem{SC3} T. Zhou and C. S. Ting, Phys.
Rev. B \textbf{80}, 224515 (2009).
%
\bibitem{SC4} Xian-Jun Zuo and Chang-De Gong,
EPL, \textbf{86} 47004 (2009).
%
\bibitem{SC5} H. Shimahara Phys. Rev. B \textbf{80},
214512 (2009).
%
\bibitem{SC6} A. Romano \emph{et al}, Phys. Rev. B \textbf{81},
064513 (2010).
%
\bibitem{SC7}  R. Ikeda, Phys. Rev. B \textbf{81}, 060510(R)
(2010).
%
\bibitem{SC8} M. M. Ma\'{s}ka \emph{et al}, Phys. Rev. B \textbf{82},
054509 (2010).
%
\bibitem{FG1} Tung-Lam Dao, A. Georges, and M. Capone, Phys. Rev. B
\textbf{76}, 104517 (2007),
%
\bibitem{FG2} Q. Chen \emph{et al}, Phys. Rev. B
\textbf{75}, 014521 (2007).
%
\bibitem{FG3} Xia-Ji Liu, H. Hu, and P. D. Drummond,
Phys. Rev. A \textbf{76}, 043605 (2007).
%
\bibitem{FG4}  M. Rizzi, \emph{et al},
Phys. Rev B \textbf{77}, 245105 (2008).
%
\bibitem{FG5} Xia-Ji Liu, Hui Hu, and P.
D. Drummond, Phys. Rev. A \textbf{78}, 023601 (2008).
%
\bibitem{FG6} M. Reza
Bakhtiari, M. J. Leskinen, and P. T\"{o}rma, Phys. Rev. Lett.
\textbf{101}, 120404 (2008),
%
\bibitem{FG7} A. Lazarides and B. Van Schaeybroec
Phys. Rev. A \textbf{77}, 041602 (2008).

%
\bibitem{FG8} X. Cui and Y. Wang, Phys.
Rev. B \textbf{79}, 180509(R) (2009).
 A. Mishra and H. Mishra, Eur.
Phys. J. D \textbf{53}, 75 (2009);.
%
\bibitem{FG9} B. Wang, Han-Dong Chen, and S.
Das Sarma, Phys. Rev. A \textbf{79}, 051604(R) (2009).
%
\bibitem{FG10} Y. Yanase,
Phys. Rev. B \textbf{80}, 220510(R) (2009).
%
\bibitem{FG11} A. Ptok, M. M\'{a}ska,
and M. Mierzejewski, J. Phys.: Condens. Matter \textbf{21}, 295601
(2009).
%
\bibitem{FG12} Yan Chen \emph{et al}, Phys. Rev. B \textbf{79}, 054512
(2009); Yen Lee Loh and N. Trivedi, Phys. Rev. Lett. \textbf{104},
165302 (2010).
%
\bibitem{FG13} A. Korolyuk, F. Massel, and P. T\"{o}rma, Phys. Rev.
Lett. \textbf{104}, 236402 (2010).
%
\bibitem{FG14} F. Heidrich-Meisner \emph{et al},
Phys. Rev. A \textbf{81}, 023629 (2010).
%
\bibitem{FG15} S. K. Baur, J. Shumway, and
E. J. Mueller, Phys. Rev. A \textbf{81}, 033628 (2010).
%
\bibitem{FG16} A. Korolyuk,
F. Massel, and P. T\"{o}rm\"{a}, Phys. Rev. Lett. \textbf{104},
236402 (2010).
%
\bibitem{FG17} M. J. Wolak \emph{et al}, Phys. Rev. A \textbf{82},
013614 (2010).
%
\bibitem{FG18} L. Radzihovsky and D. Sheehy, Rep. Prog. Phys.
\textbf{73},  076501 (2010)
%
\bibitem{HTa} J. M. Edge and N. R. Cooper, Phys. Rev. Lett. \textbf{103}, 065301 (2009);
 Phys. Rev. A
\textbf{81}, 063606 (2010).
%
\bibitem{DMa}  A. Sedrakian and D. H.
Rischke, Phys. Rev. D \textbf{80}, 074022 (2009).
%
\bibitem{Sa} G. Sarma, J. Phys. Chem. \textbf{24}, 1029 (1963).
%
\bibitem{FD1} P. F. Bedaque, H. Caldas, and G. Kupak, Phys. Rev. Lett.
\textbf{91}, 247002 (2003).
%
\bibitem{FD2} H. Caldas, Phys Rev. A \textbf{69}, 063602
(2004).
%
\bibitem{FD3} H. Caldas, C. W. Morais and A. L. Mota, Phys. Rev. D \textbf{ 72},
045008 (2005).
%
\bibitem{FD4} S. Sachdev and K. Yang, Phys. Rev. B \textbf{73}, 174504
(2006).
%
\bibitem{FDFS1} D.E. Sheehy, L. Radzihovsky, Phys. Rev. Lett. \textbf{96}, 060401 (2006).
%
\bibitem{FDFS2}  L. He, M. Jin, and  P. Zhuang, Phys. Rev. B \textbf{74}, 024516 (2006).
%
\bibitem{FDFS3} D.E. Sheehy and L. Radzihovsky, Ann. Phys. (N.Y.) \textbf{322}, 1790 (2007).
%
\bibitem{FDFS4} M. Iskin and C.A.R. Sá de Melo, Phys. Rev. A \textbf{78}, 013607 (2008).
%
\bibitem{SZ} S. Palh, and Z.  Koinov, J Low. Temp. Phys., \textbf{176}, 113 (2014)

\end{thebibliography}
\end{document}